\documentclass[cqg,10pt]{iopart}
\usepackage{iopams}
 \expandafter\let\csname equation*\endcsname\relax
  \expandafter\let\csname endequation*\endcsname\relax
\usepackage{amssymb, amsmath, graphicx}
\usepackage[centering,hmargin=2cm,vmargin=2.5cm]{geometry}
\usepackage[utf8]{inputenc}
\usepackage{soul}
\usepackage[titletoc,title]{appendix}
\usepackage{subfloat}
\usepackage{color}

\newcommand{\bra}[1]{\langle#1 \vert}
\newcommand{\ket}[1]{ \vert #1\rangle}

\newcommand{\uu}{\uparrow}
\newcommand{\dd}{\downarrow}

\newenvironment{psmallmatrix}
  {\left(\begin{smallmatrix}}
  {\end{smallmatrix}\right)}

\begin{document}
\date{\today}
\title[Quantum-enhanced multi-parameter estimation for unitary photonic systems]{Quantum-enhanced multi-parameter estimation for unitary photonic systems}

\author{Nana Liu$^{1,2,3}$, Hugo Cable$^4$}

\address{$^1$ Clarendon Laboratory, Department of Physics, University of Oxford, Oxford OX1 3PU, United Kingdom}
\address{$^2$ Singapore University of Technology and Design, 8 Somapah Road, Singapore 487372}
\address{$^3$ Centre for Quantum Technologies, National University of Singapore, 3 Science Drive 2, Singapore 117543}
\address{$^4$ Quantum Engineering Technology Labs, H. H. Wills Physics Laboratory and Department of Electrical and Electronic Engineering, University of Bristol, BS8 1FD, United Kingdom}
\ead{Nana.Liu@physics.ox.ac.uk; Nana.Liu@quantumlah.org}
\ead{Hugo.Cable@bristol.ac.uk}

\begin{abstract}
Precise device characterization is a fundamental requirement for a large range of applications using photonic hardware, and constitutes a multi-parameter estimation problem.   Estimates based on measurements using single photons or classical light have precision which is limited by shot-noise, while quantum resources can be used to achieve sub-shot-noise precision.  However, there are many open questions with regard to the best quantum protocols for multi-parameter estimation, including the ultimate limits to achievable precision, as well as optimal choices for probe states and measurements.  In this paper, we develop a formalism based on Fisher information to tackle these questions for set-ups based on linear-optical components and photon-counting measurements. A key ingredient of our analysis is a mapping for equivalent protocols defined for photonic and spin systems, which allows us to draw upon results in the literature for general finite-dimensional systems. Motivated by the protocol in X.-Q. Zhou, et al., Optica 2, 510 (2015), we present new results for quantum-enhanced tomography of unitary processes, including a comparison of Holland-Burnett and NOON probe states.
\end{abstract}

\maketitle

\section{Introduction}

Advances in precision measurement are playing an ever more important
role in technological development. From biological imaging \cite{bioimaging}, quantum clocks \cite{quantumclock, bloom2014optical} quantum computing \cite{o2007optical}, thermometry \cite{toyli2013fluorescence}, to the recent
detection of gravitational waves \cite{LIGO}, there is an increasing demand for higher
precision in parameter estimation schemes using light. Quantum resources have been shown to serve a crucial role in pushing beyond the precision limits available to classical probes, especially in single-parameter estimation \cite{giovannetti2011advances}. The quantum advantage in multi-parameter estimation is less well studied. However, applications like imaging, linear-optical quantum computing and characterisation of optical fibres for quantum communication require the simultaneous estimation of multiple parameters. Thus, a question of chief interest for future applications is to find the quantum resources for multi-parameter problems that offer optimal precision while remaining experimentally accessible.  

One primary resource in parameter estimation is particle number $N$. Parameter estimation using classical resources is constrained by the shot-noise limit \cite{giovannetti2011advances}, for which precision (mean-square error) scales as $1/N$ \cite{giovannetti2011advances}. This is true in conventional multi-parameter estimation using single-photon probes (process tomography) \cite{nandc}. There are situations however, such as when probing delicate samples, where the use of high $N$ is undesirable \cite{bioimaging}.

It is well-known for the idealised case, when the effects of particle losses and decoherence can be ignored, that quantum resources enable up to quadratic improvement in precision for both single parameter estimation and multi-parameter estimation. This is called Heisenberg scaling, i.e., precision scales like $\mathcal{O}(1/N^2)$ \cite{Bollinger1996}. In single-parameter estimation, NOON states uniquely achieve the optimal precision (Heisenberg limit) using photon-number-counting measurements when again losses and decoherence are neglected \cite{durkin2007}. In contrast, for the task of estimating a set of parameters that fully characterize any linear-optical process, it has not previously been shown what conditions must be satisfied to achieve optimal precision, or even what this optimal precision is.

One of the first experimental demonstrations of Heisenberg scaling for a general linear-optical process was recently performed \cite{hugotomo} (for $N=4$ ). It was based on a new protocol for characterising an unknown two-mode linear-optical process, using Holland-Burnett states \cite{HB1993} and photon-number-counting measurements. This is equivalent to estimating the independent parameters of an unknown $SU(2)$ matrix. The analysis of \cite{hugotomo} uses process fidelity to quantify precision, as is typical for process tomography. However, using process fidelity, it is difficult to establish the optimality of the protocol, which is essential to enable comparisons between alternative choices of probe states and measurement schemes.

A theoretical tool very well-suited to this analysis is the Fisher information formalism, which is already much-used in estimation theory \cite{giovannetti2011advances, liero2016introduction}. For photonic systems, Fisher information has already been exploited for some very specific cases of multi-parameter estimation where substantial simplifications occur. For example, when parameters are associated with commuting operations \cite{ciampini, peterh, yue} or single-parameter estimation with environment interaction \cite{mihai, philip}. There has been active research on spin systems in related contexts. A succession of theoretical studies, using fidelity measures to characterise precision \cite{bagan1, bagan2, chiri, hayashi} and using Fisher information \cite{ballester2004, ballester2005, animesh} have demonstrated how Heisenberg scaling can be achieved. Theoretical studies of $SU(2)$-estimation using the Fisher information matrix have provided mathematical conditionals for achieving optimal precision \cite{fujiwara, matsumoto}. 

In this paper, we develop theoretical machinery to explore $SU(2)$-estimation protocols for linear-optical set-ups using quantum states. We provide simple conditions to test which photonic states are optimal (extending results in \cite{ballester2005} for spin systems). We also interpret these results in terms of optical interferometry. As two important examples, we show that both Holland-Burnett states and NOON states are optimal, but we find that neither are optimal using photon-number-counting measurements. Furthermore, we demonstrate that the precision in this context is dependent on the unitary itself. In addition to having important practical implications, these results show that multi-parameter estimation cannot be considered a simple generalisation of single-parameter estimation.  

We also introduce a mapping between photonic and spin states that allow us to translate the results for photonic systems to spin systems and vice-versa. This analogy allows us to compare the spin analogue of our optical protocol to those of existing multi-parameter estimation schemes using spin systems. For instance, this makes it possible to prove that a probe state used in \cite{animesh} is indeed optimal, which was suspected but not proved. We also show that the spin analogue of our protocol contains a larger class of optimal states compared to \cite{animesh}. 

After establishing our mapping from photonic to spin states and processes in Sec.~\ref{sec:photonmap}, we turn to a  brief introduction to the experimental protocol in \cite{hugotomo} in Sec.~\ref{sec:protocol}. We then introduce the basics of the Fisher information formalism in Sec.~\ref{sec:FIsection}. In Sec.~\ref{sec:Optimal precision and states} we extend previous results in the literature to find the optimal precision and the conditions for optimal states in our protocol. We also discuss the implications of these results for spin systems. In Sec.~\ref{sec:application}, we apply our results to study the special cases of Holland-Burnett states and NOON states under photon-number-counting measurements before summarising our main results and future directions in Sec.~\ref{sec:summary}. 

\section{Background}

\subsection{Equivalent protocols for photonic and spin systems}
\label{sec:photonmap}
We now establish a mapping between an $N$-particle two-mode linear-optical process and a process with $N$ spin-$1/2$ particles. This is important for making a formal analogy for protocols specified for photonic and spin systems. It can be used for translating results on multi-parameter estimation using spin systems to the context of photonic multi-parameter estimation. For standard mappings between bosonic and spin states and operators, see \cite{holstein1940field, biedenharn1965quantum}.\\

We begin with an $N$-particle two-mode photonic state $\ket{M, N-M}$, where $M$ is an integer $0 \leq M \leq N$. There is a one-to-one correspondence between this state and an $N$-particle spin-$1/2$ state that remains invariant with respect to any particle exchange (i.e., symmetric) 
\begin{align} \label{eq:mapping}
\ket{M, N-M}  \longleftrightarrow \frac{1}{\sqrt{\binom{N}{M}}}\sum_j \Pi_j \left(\ket{\uu}^{\otimes M} \otimes \ket{\dd}^{\otimes (N-M)}\right) \equiv \ket{\xi_0}_{\text{spin}}.
\end{align}
The summation $\sum_j \Pi_j$ is over all the possible permutations of the product states and $\ket{\xi_0}_{\text{spin}}$ is also known as a Dicke state \cite{Dicke}. For concreteness, we choose $\ket{\uu}$, $\ket{\dd}$ to be the spin-up and spin-down eigenstates of Pauli matrix $\sigma_z$. We denote the creation operators for the two photonic modes by $a^{\dagger}$ and $b^{\dagger}$, obeying commutation relations $[a,a^{\dagger}]=[b,b^{\dagger}]=\mathbf{1}$. The creation operators corresponding to the up and down spin states are represented by $a^{\dagger}_{\uu}$ and $a^{\dagger}_{\dd}$. These satisfy the anticommutation relations $\{a^{\dagger}_{\uu},a_{\uu}\}=\mathbf{1}=\{a^{\dagger}_{\dd},a_{\dd}\}$ and where all other anticommutation relations vanish. In the single-particle case, we can make the correspondence $a^{\dagger} \ket{0,0} \longleftrightarrow a^{\dagger}_{\uu} \ket{0}=\ket{\uu}$ and $b^{\dagger} \ket{0,0}\longleftrightarrow a^{\dagger}_{\dd}\ket{0}=\ket{\dd}$. Extending this to $N$ particles gives the mapping between photonic states and spin states that we require. 
\begin{figure}[ht!]
\centering
\includegraphics[scale=0.3]{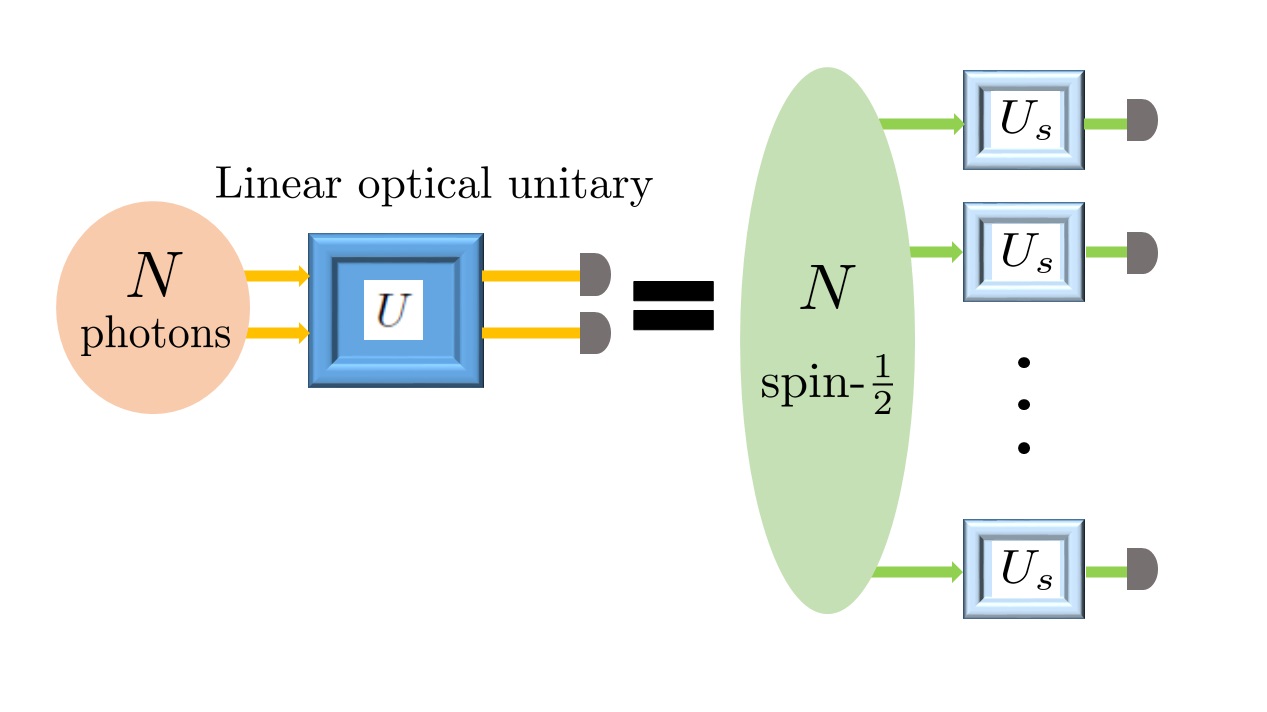}
\caption[\textit{Linear-optical process and corresponding evolution of spin system}.]{\label{operatormap}\textit{Linear-optical process and corresponding evolution of spin system}. Mapping between a two-mode linear-optical process $U$ acting on $N$ photons, and the analogous process with $N$ spin-$1/2$ particles each undergoing evolution $U_s$. Photonic superposition states correspond to superpositions of spin states that are symmetric under particle exchange. Measurement of $M'$ (or $N-M'$) photons in the first (or second) mode correspond to measurement of $M'$ spin-up (or $N-M'$ spin-down) particles in the spin picture (where the ordering is ignored).}
\end{figure}

This mapping also allows us to describe a transformation of the two-mode photonic state under unitary operator $U$ in terms of the evolution of $N$ spin-$1/2$ particles, described by the unitary operator $U_s$ (see Fig.~\ref{operatormap}). It can be shown that the following correspondence holds (for a derivation see Appendix~\ref{app:photonmap})
\begin{align}
U\ket{M, N-M} \longleftrightarrow U_s^{\otimes N} \ket{\xi_0}_{\text{spin}}.
\end{align}
Each two-mode linear-optical unitary corresponds to a two-by-two unitary matrix.  In the remainder of this paper, we will disregard any global phase for $U$, which can then be characterized by three parameters, and represented by an $SU(2)$ matrix. 
 
We can complete the analogy by mapping photon-number-counting measurement to measurements of the spin system. We note that a projective measurement $\ket{M',N-M'}\bra{M',N-M'}$ on the photonic state (where $M'$ is an integer $0\leq M'\leq N$) corresponds in the spin picture to measurement of $M'$ spin-up particles and $N-M'$ spin-down particles, where there are $N!/(M'!(N-M')!)$ equivalent measurement patterns. 

We can use this mapping to show that the $N$-particle NOON states $(\ket{N,0}+\ket{0,N})/\sqrt{2}$ map to $N$-particle GHZ \cite{greenberger1989going} states $(\ket{0}^{\otimes N}+\ket{1}^{\otimes N})/\sqrt{2}$. This explains why both NOON states and GHZ states have been found to be optimal states (i.e., achieving optimal precision) in single-parameter estimation \cite{toth}, despite being used in different types of set-ups. Another example is the correspondence between $N$-particle Holland-Burnett states $\ket{N/2, N/2}$ and symmetric Dicke states with $N/2$ excitations \cite{toth}.  
\subsection{Quantum-enhanced protocol for unitary estimation}
\label{sec:protocol}
To identify an unknown optical process, process tomography is traditionally used. It relies on single-photon probes or classical light and is shot-noise limited. However, when using non-classical multi-photon probe states, much greater precision per number of photons can be achieved. In this section, we briefly describe the multi-photon probe scheme for $SU(2)$-estimation recently performed in \cite{hugotomo}.

\begin{figure}[ht!]
\centering
\includegraphics[scale=0.4]{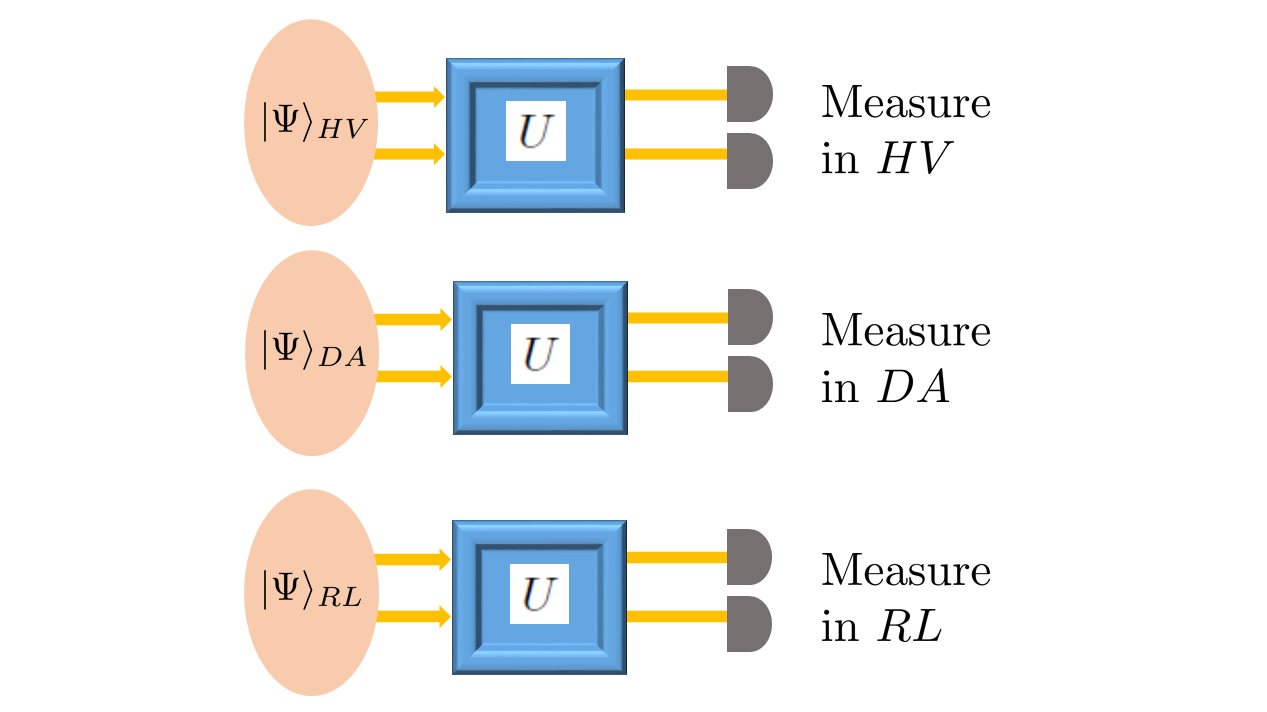}
\caption[\textit{A \textit{single run} of the multi-parameter
estimation protocol}.]{\label{protocol}\textit{A \textit{single run} of the multi-parameter
estimation protocol investigated in \cite{hugotomo}.} Begin with three input states $\ket{\Psi}_{HV}$, $\ket{\Psi}_{DA}$, $\ket{\Psi}_{RL}$. These are passed through the unitary $U$ before being respectively measured in bases $HV$, $DA$ and $RL$.}
\end{figure}

We begin with a two-mode $N$-particle photonic state $\ket{\Psi}_{HV}=\ket{M, N-M}_{HV}$ in the $HV$ (horizontal and vertical) polarisation basis. This state is passed through a linear-optical process, which can be characterised by a $SU(2)$ matrix. It is possible to recover the probability distributions of photon numbers in each of the two modes after passing through the unknown unitary by using photon-number-counting measurements with respect to the $HV$ basis. This procedure can be repeated with respect to the $DA$ (diagonal and anti-diagonal) and $RL$ (right and left-circular) polarisation bases. By definition $\ket{1,0}_{DA}=(1/\sqrt{2})(\ket{1,0}_{HV}+\ket{0,1}_{HV})$, $\ket{0,1}_{DA}=(1/\sqrt{2})(\ket{1,0}_{HV}-\ket{0,1}_{HV})$ and $\ket{1,0}_{RL}=(1/\sqrt{2})(\ket{1,0}_{HV}+i\ket{0,1}_{HV})$, $\ket{0,1}_{RL}=(1/\sqrt{2})(\ket{1,0}_{HV}-i\ket{0,1}_{HV})$. A \textit{single run} of the $SU(2)$-estimation protocol is defined as the procedure above performed with respect to all three $HV$, $DA$ and $RL$ polarisation bases. The three independent parameters of the unknown $SU(2)$ matrix can then be fully recovered from the set of three photon number probability distributions (see \cite{hugotomo}). See Fig.~\ref{protocol} for a diagram representing a single run of this protocol.

When this protocol is performed $\nu$ times, the total number $n$ of input photons used is thus $3N \nu$. It can be experimentally shown \cite{hugotomo} that, for up to $N=4$, the precision for estimating each of a full set of parameters which characterize the unknown unitary using Holland-Burnett input states $\ket{\Psi}=\ket{N/2, N/2}$ is $\mathcal{O}(1/(\nu N^2))$. This is Heisenberg scaling. This is contrasted with the shot-noise limited precision using single-photon probes, which has scaling $\mathcal{O}(1/(\nu N))$. In \cite{hugotomo}, the performance of the protocol for different choices of unitary and probe states was compared using process fidelity. 

\begin{figure}[ht!]
\centering
\includegraphics[scale=0.35]{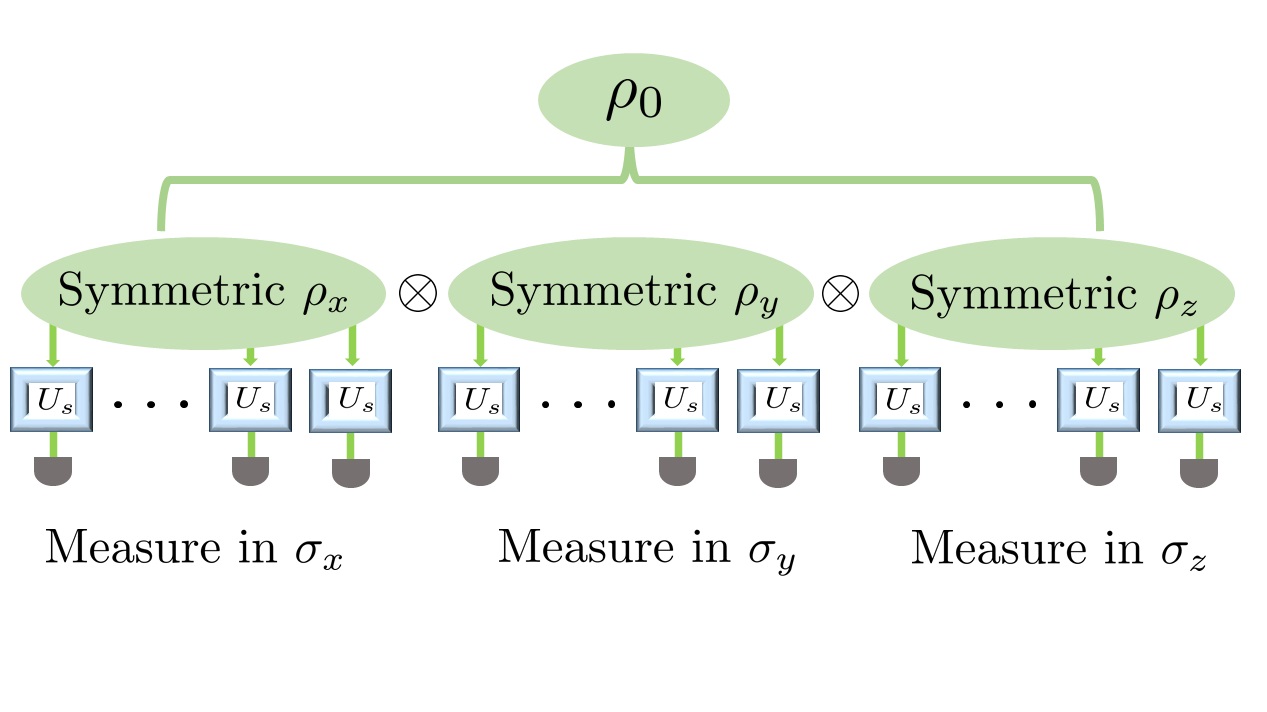}
\caption{\label{fig:spinprocess}\textit{A \textit{single run} of the spin analogue to the linear-optical multi-parameter
estimation protocol.} Begin with the input symmetric spin states $\rho_x$, $\rho_y$ and $\rho_z$, where each particle is passed through a unitary operation $U_s$. Then measure in bases $\sigma_x$, $\sigma_y$ and $\sigma_z$ respectively.}
\end{figure}

We note that the HV, DA and RL polarisation bases respectively map to the $z$, $x$ and $y$ bases in the spin case. We denote here $\rho_x$, $\rho_y$ and $\rho_z$ as the symmetric spin-state analogues to the photonic input states in bases $HV$, $DA$ and $RL$. The probe state $\rho_0$ for a single run of the analogous spin system protocol (i.e., $\nu=1$) is of the form $\rho_0=\bigotimes_{k=x, y, z} \rho_k$. See Fig.~\ref{fig:spinprocess} for a spin-system analogue of the linear-optical unitary-estimation protocol. 
\subsection{Fisher information}
\label{sec:FIsection}
\subsubsection{Cram\'er-Rao inequality for single and multi-parameter estimation}
In any scheme to estimate unknown parameters, it is useful to
bound the variance of those parameters as a way of characterising the precision. The inverse of a quantity known as the Fisher information provides a means to bound variances in local parameter estimation, i.e., beginning from a 
rough estimate of the parameters and making this estimate more precise. 
This type of bound is provided by the Cram\'er-Rao inequality. It is
this bound that makes Fisher information
so crucial and the reason we choose this measure to characterise precision. 

In any given process, the goal is to find the initial state
and measurement maximising the Fisher information while being subject to some
given constraints of one's resources. Here we take this resource to be the total number of particles used in probe states over all runs of the estimation protocol.

To estimate a single unknown parameter $\theta$, the Cram\'er-Rao inequality states \cite{cramer1947mathematical} that
the variance of the unbiased estimators of $\theta$ achievable is 
\begin{align}
\delta \theta^2 \geq \frac{1}{F_{\theta}}.
\end{align}
By an unbiased estimator we mean that the average value is exactly $\theta$. The Fisher information is denoted $F_{\theta}$ and is defined by
\begin{align}
F_{\theta}=\sum_{M_d} \frac{1}{P(M_d;\theta)}\left(\frac{\partial P(M_d;\theta)}{\partial \theta}\right)^2.
\end{align}
The probability the final state of the protocol $\rho_{\theta}$ has the (detected) measurement outcome $M_d$ is denoted $P(M_d;\theta)$.

This bound can be made tighter by introducting quantum Fisher information $I_{\theta}$. This 
is defined as the maximum Fisher information with respect to all possible final measurements. It can be shown that
$F_{\theta} \leq I_{\theta}=\tr(\rho_{\theta} \lambda_{\theta}^2)$, where $\lambda_{\theta}$ is the symmetric logarithmic derivative (SLD) \cite{braunsteincaves} and is 
defined by $\partial_{\theta} \rho_{\theta}=(1/2)(\lambda_{\theta} \rho_{\theta}+\rho_{\theta}\lambda_{\theta})$. This introduces the quantum Cram\'er-Rao inequality
\begin{align}
\delta \theta^2 \geq \frac{1}{F_{\theta}} \geq \frac{1}{I_{\theta}},
\end{align}
where the equality can always be saturated and the optimal precision is achievable by known initial states and final measurements \cite{braunsteincaves}. 

The optimal achievable precision from the quantum Cram\'er-Rao  inequality is $\delta \theta^2=1/(\nu N^2)$. This is uniquely achieved using the $N$-particle NOON state.
The optimal measurements, which include photon-number-counting measurements, can be expressed in terms of the eigenvectors of the SLD. Another state that can achieve
Heisenberg scaling (i.e., $\delta \theta^2=\mathcal{O}(1/ (\nu N^2))$), but with sub-optimal precision, is the Holland-Burnett state \cite{durkin2007}. Other sub-optimal states with Heisenberg
scaling include Yurke states \cite{rosetta, yurke1986input, yurke19862}, amongst many others \cite{durkin2007, rosetta}. 

In the estimation of \textit{multiple} parameters $\{ \theta_{\alpha}\}$, precision is captured by the covariance matrix $C$, which is defined by
$C_{\alpha \beta} \equiv \langle \theta_{\alpha} \theta_{\beta}\rangle-\langle \theta_{\alpha}\rangle \langle \theta_{\beta}\rangle$ and $\langle \cdot \rangle$ denotes an average over all measurement outcomes. Note that the covariance matrix reduces to 
the variance in the case of a single parameter. The covariance matrix can be bounded by the inverse of the Fisher information matrix $F$, defined by
\begin{align}\label{eq:FImatrix}
F_{\alpha \beta}= \sum_{M_d} (1/P(M_d; \{ \theta_{\gamma}\})) 
(\partial P(M_d; \{ \theta_{\gamma}\})/\partial \theta_\alpha)(\partial P(M_d; \{ \theta_{\gamma}\})/\partial \theta_\beta).
\end{align} 
$F_{\alpha \beta}$ is in turn upper bounded by the quantum Fisher information matrix $I$, which defined by \cite{helstrom1967minimum, helstrom1968minimum, parisreview}
\begin{align} \label{eq:QFIdef}
I_{\alpha \beta}=\text{Re}\tr(\rho_{\theta}\lambda_{\alpha} \lambda_{\beta}),
\end{align} 
where for pure states $\lambda_{\alpha}=2 \partial \rho_{\theta}/\partial \theta_\alpha$. A multi-parameter quantum Cram\'er-Rao  inequality can thus be written as 
\begin{align} \label{eq:crmatrix}
\tr(C) \geq \tr(F^{-1}) \geq \tr(I^{-1}).
\end{align}
This follows from the inequalities $C\succcurlyeq F^{-1} \succcurlyeq I^{-1}$, where $C-I^{-1} \succcurlyeq 0$ denotes $C-I^{-1} $ is a positive semi-definite matrix \cite{helstrom1967minimum, helstrom1968minimum, parisreview}. When the equality in $C-I^{-1} \succcurlyeq 0$ is satisfied, this is known as saturating the multi-parameter quantum Cram\'er-Rao inequality and implies $\tr(C)=\tr(F^{-1})=\tr(I^{-1})$. 

However, unlike in single-parameter estimation, it is not always possible to always saturate the multi-parameter quantum Cram\'er-Rao inequality. This is because the operators corresponding to optimal measurements (constructed from the eigenvectors of $\lambda_{\alpha}$) corresponding to different parameters may not commute. Hence it is not possible to implement them simultaneously and thus optimize precision with respect to all the parameters. However, there are cases where saturation is possible and for pure states this is investigated in Sec.~\ref{sec:Optimal precision and states}. For a discussion of differences between cases with pure and mixed states, see \cite{ragy2016compatibility}.

For recent reviews on this subject, we refer interested readers to \cite{parisreview, toth, szczykulska2016multi} and references therein. 
\subsubsection{Quantum Fisher information matrix}
\label{sec:QFIoriginalderivation}
Now we derive the general form for protocols using pure probe states. We consider a general pure probe state $\rho_0$, unitary transformation $\tilde{U}$, and final state $\rho_{\theta}=\tilde{U} \rho_0 \tilde{U}^{\dagger}$. It is convenient to define $\tau_{\alpha}=i\tilde{U}^{\dagger} \partial \tilde{U}/\partial \theta_{\alpha}$, which is the generator of $\tilde{U}$ and which must be Hermitian. Then the quantum Fisher information matrix is given by
\begin{align}\label{eq:QFIdef2}
I_{\alpha \beta}(\rho_0)=4 \text{Re}\left(\text{tr} \left(\rho_{\theta}\frac{\partial \rho_{\theta}}{\partial \theta_{\alpha}}\frac{\partial \rho_{\theta}}{\partial \theta_{\beta}}\right)\right). 
\end{align}
Using the unitarity of $\tilde{U}$, the cyclic-permutation invariance of the trace and purity $\rho_0^2=\rho_0$, it is straightforward to show  $\tr(\rho_{\theta}\lambda_{\alpha} \lambda_{\beta})=4[\tr(\rho_0\tau_{\alpha}\tau_{\beta})-\tr(\rho_0 \tau_{\alpha}\rho_0 \tau_{\beta})]$. Since $\rho_0$ is again pure, $\tr(\rho_0 \tau_{\alpha}\rho_0 \tau_{\beta})]=\tr(\rho_0 \tau_{\alpha})\tr(\rho_0 \tau_{\beta})$ and
\begin{align}
I_{\alpha \beta}(\rho_0)=4 \text{Re}[\tr(\rho_0\tau_{\alpha}\tau_{\beta})]-4\tr(\rho_0 \tau_{\alpha})\tr(\rho_0 \tau_{\beta}).
\end{align}
We turn now to the spin protocol considered in this paper, where we saw $\rho_0$ is an $n$-particle pure state undergoing unitary transformation $\tilde{U}=U_s^{\otimes n}$. Imposing no other restrictions on $\rho_0$ other than purity, we want to express $I_{\alpha \beta}$ in terms of the generators of $U_s$, which are $t_{\alpha} \equiv iU_s^{\dagger} \partial U_s/\partial \theta_\alpha$. We observe the relations
$\tau_{\alpha}=t_{\alpha}\otimes \mathbf{1}^{\otimes (n-1)}+\mathbf{1}\otimes t_{\alpha} \otimes \mathbf{1}^{\otimes (n-2)}+...+\mathbf{1}^{\otimes (n-1)} \otimes t_{\alpha}$ (which has $n$ terms altogether) and $\tau_{\alpha} \tau_{\beta}=t_{\alpha}t_{\beta}\otimes \mathbf{1}^{\otimes (n-1)}+t_{\alpha}\otimes t_{\beta}\otimes \mathbf{1}^{\otimes (n-2)}+t_{\alpha}\otimes\mathbf{1}^{\otimes (n-1)}\otimes t_{\beta}+...+\mathbf{1}^{\otimes (n-1)}\otimes t_{\alpha}t_{\beta}$ (which has $n^2$ terms altogether). These relations together with standard trace equalities yield
\begin{align} \label{eq:totalqfisherinfo}
&I_{\alpha \beta}(\rho_0)= 4\text{Re}\sum_{i=1}^{n} \tr(\tr_{[i]}(\rho_0) t_{\alpha}t_{\beta}) \nonumber \\
&+4\sum_{i \neq j=1}^{n}\tr(\tr_{[i,j]}(\rho_0) (t_{\alpha}\otimes t_{\beta})) \nonumber \\
&-4\left(\sum_{i=1}^{n} \tr(\tr_{[i]}(\rho_0) t_{\alpha})\right)\left(\sum_{j=1}^{n} \tr(\tr_{[j]}(\rho_0) t_{\beta})\right).
\end{align}
Here the notation $\tr_{[i,j]}$ denotes a trace over all particles \textit{except} those particles labelled by positions $i,j$. The ordering of $i,j$ matters so that the reduced state $\tr_{[j,i]}(\rho_0)$ corresponds to $\tr_{[i,j]}(\rho_0)$ with a swap of the two subsystems $i$ and $j$. Interestingly, this means that the total Fisher information is dependent only on the one-particle and two-particle reduced density matrices of $\rho_0$. 

Our Eq.~\eqref{eq:totalqfisherinfo} holds for any unitary $U_s$. In a special instance when $U_s=\exp(iH)$, where $H=\sum_{\alpha} \theta_{\alpha} h_{\alpha}$ and $h_{\alpha}$ is independent of unknown parameters $\theta_{\alpha}$, Eq.~\eqref{eq:totalqfisherinfo} agrees with \cite{animesh}. We also note that \cite{ballester2005} considers a closely-related case where the protocol incorporates an ancilla and $\rho_0$ is symmetric.  
\subsubsection{Parameterisation}
In Sec.~\ref{sec:photonmap}, we saw that the unitary operator $U_s$ is sufficient to describe a two-mode linear-optical process and can be represented by a $SU(2)$ matrix. We now need to define a parameterisation of this $SU(2)$ matrix. This helps us compute the quantum Fisher information and Fisher information matrices in a way which allows us to meaningfully quantify the corresponding notions of statistical information about $U_s$. To this end, we can express this parameterisation as a restriction on the generators $\{t_{\alpha}\}$ in the following way. 

Since we are performing a local parameter estimation of $U_s$, we can define $U_s$ in terms of a local expansion about some known unitary operator $U_s^{(0)}$. The higher precision to which this expansion is known, the better estimate we have of $U_s$. $U_s$ is dependent only on the three \textit{unknown} parameters $\{\theta_{\alpha}\}$. Let $U_s^{(0)}$ be defined only in terms of the three \textit{known} parameters $\{\theta_{\alpha}^{(0)}\}$. To linear order, a Taylor expansion about $U_s^{(0)}$ can be written as $U_s(\{\theta_{\alpha}\}) \equiv U_s^{(0)}(\{\theta_{\alpha}^{(0)}\})(\mathbf{1}-i\sum_m t_{\alpha}\vert_{\theta_{\alpha}=\theta_{\alpha}^{(0)}}(\theta_{\alpha}-\theta_{\alpha}^{(0)}))$. 

One good parameterisation is where precision is independent of the particular $U_s^{(0)}$, i.e., the value of the initial guess for $U_s$. This is achieved when $t_{\alpha}\vert_{\theta_{\alpha}=\theta_{\alpha}^{(0)}}$ is independent of $\{\theta_{\alpha}^{(0)}\}$, which means $t_{\alpha}$ is independent of $\{\theta_{\alpha}\}$. A natural decision is choosing $\{t_{\alpha}\}$ to be proportional to the Pauli spin matrices $\{\sigma_{\alpha}\}$ (we note that this parameterisation was also adopted in \cite{ballester2005} and provides simplification of the analysis). We will call parameters $\{\theta_{\alpha}\}$ satisfying $t_{\alpha}=\sigma_{\alpha}/\sqrt{2}$ \textit{locally-independent parameters}. We adopt this terminology since in every local region about some $U_s^{(0)}$, the generators $\{t_{\alpha}\}$ are independent of $U_s^{(0)}$. This is not true in more general parameterisations, like the Euler angle parameterisation.

\section{Optimal precision and conditions for optimal states}
\label{sec:Optimal precision and states}

In this section we demonstrate the necessary and sufficient conditions our photonic-probe states must satisfy to reach optimal precision and we find what this optimal precision is. Our results also allow us to readily identify these photonic states.

Reaching the optimal precision allowed by the quantum Cram\'er-Rao inequality consists of two parts: saturating the quantum Cram\'er-Rao  inequality and attaining the smallest value of the trace of the inverse quantum Fisher information matrix. We examine these conditions and their implications in Sec.~\ref{sec:SaturateCR} and Sec.~\ref{sec:optimalproof} respectively. 

We proceed by considering protocols for spin systems and deriving optimality results along similar lines to \cite{ballester2005} for ancilla-based protocols. Then utilising our mapping introduced in Sec.~\ref{sec:photonmap}, we translate these results in terms of photonic states and processes. We find they lead to novel and important consequences for optical metrology which we explore.

We also show how the spin-system analogue to our photonic unitary-estimation protocol allows for a much wider class of optimal states than has been identified \cite{animesh}.
\subsection{Quantum Fisher information matrix for unitary estimation}
The spin-system analogue to our unitary-estimation protocol is represented in Fig.~\ref{fig:spinprocess}. It begins with a $n$-particle input state $\rho_0=\bigotimes_{k=x, y, z}\rho_k$, where $\rho_k$ are $N$-particle input states. For a single run of the protocol, $I_{\alpha \beta}$ is a sum of the quantum Fisher information matrices with respect to each sub-protocol. Thus
\begin{align}\label{eq:qfiadditivity}
I_{\alpha \beta}(\rho_0)=\sum_{k=x,y,z} I_{\alpha \beta}(\rho_k).
\end{align}
This additivity condition follows from Eq.~\eqref{eq:QFIdef2}, the purity of $\rho_0$ and $\tilde{U}=U_s^{\otimes 3N}$.

The state $\rho_0$ is in general not a symmetric state, while each $\rho_k$ is symmetric (see Sec.~\ref{sec:photonmap}). This means all partial traces of $\rho_k$ depend only on the number of subsystems traced out and not on \textit{which} subsystem is traced out. Therefore all one-particle reduced states of $\rho_k$ are identical and we can define $\rho^{[1]}_k \equiv \tr_{[i]}(\rho_k)$ for any $i$. All two-particle reduced states are also identical and we define $\rho^{[2]}_k \equiv \tr_{[i,j]}(\rho_k)$ for any $i,j$ where $i \neq j$. Applying our locally-independent parameterisation $t_{\alpha} \equiv \sigma_{\alpha}/\sqrt{2}$ in Eqs.~\eqref{eq:totalqfisherinfo} and ~\eqref{eq:qfiadditivity}, the quantum Fisher information matrix for our protocol simplifies to the form
\begin{align}\label{eq:newQFIfull}
I_{\alpha \beta}(\rho_0)=\text{Re}\{2N[\tr(\rho^{[1]}_{\text{tot}}\sigma_{\alpha}\sigma_{\beta})] \}+2N(N-1)\tr(\rho^{[2]}_{\text{tot}}(\sigma_{\alpha}\otimes \sigma_{\beta}))-2N^2\sum_{k=x,y,z}\tr(\rho^{[1]}_k\sigma_{\alpha}) \tr(\rho^{[1]}_k\sigma_{\beta}),
\end{align}
where $\rho_{\text{tot}}^{[1]} \equiv \sum_{k=x, y, z} \rho_k^{[1]}$ and $\rho_{\text{tot}}^{[2]} \equiv \sum_{k=x, y, z} \rho_k^{[2]}$. 
\subsection{Saturating the quantum Cram\'er-Rao inequality}
\label{sec:SaturateCR}
Saturation of the quantum Cram\'er-Rao inequality is a non-trivial constraint in the multi-parameter setting. For pure states, it can be shown that a necessary and sufficient condition to attain this saturation is to satisfy $\text{Im} [\tr(\rho_{\theta} \lambda_{\alpha} \lambda_{\beta})]=0$ \cite{matsumoto}. Since $\lambda_{\alpha}$ are hermitian, this condition is equivalent to 
\begin{align}\label{eq:lambdacommrelation}
\tr(\rho_{\theta} [\lambda_{\alpha}, \lambda_{\beta}])=0.
\end{align}
Following similar arguments as in Sec.~\ref{sec:QFIoriginalderivation}, for arbitrary $\tilde{U}$ and pure states $\rho_{\theta}=\tilde{U}\rho_0\tilde{U}^{\dagger}$, Eq.~\eqref{eq:lambdacommrelation} is equivalent to
\begin{align} \label{eq:taualphataubeta}
\tr(\rho_0 [\tau_{\alpha}, \tau_{\beta}])=0.
\end{align}
For the scenario pictured in Fig.~\ref{fig:spinprocess} with $\tilde{U}=U^{\otimes 3N}$ we now obtain $\tr(\rho_0 \tau_{\alpha} \tau_{\beta}) 
=\sum_{i=1}^{3N}\tr[\tr_{[i]}(\rho_0)(t_{\alpha}t_{\beta})] 
+\sum_{i \neq j=1}^{3N}\tr[\tr_{[i,j]}(\rho_0)(t_{\alpha}\otimes t_{\beta})]$. Using $\rho_0=\bigotimes_{k=x, y, z} \rho_k$, Eq.~\eqref{eq:taualphataubeta} thus reduces to
\begin{align} 
\tr(\rho^{[1]}_{\text{tot}}[t_{\alpha},t_{\beta}])=0.
\end{align}
For the locally-independent parameterisation this becomes $\tr(\rho^{[1]}_{\text{tot}}[\sigma_{\alpha}, \sigma_{\beta}])=0$, which is uniquely satisfied when $\rho^{[1]}_{\text{tot}} \propto \mathbf{1}$. We can write $\rho_z^{[1]}=\mathbf{1}/2+\sum_{k=x, y, z} b_k \sigma_k$ where $b_{k}$ are constants. Then using $\rho^{[1]}_x=h \rho^{[1]}_z h^{\dagger}$ and $\rho^{[1]}_y=h_c \rho^{[1]}_z h_c^{\dagger}$, where  $h=(1/\sqrt{2})\begin{psmallmatrix}
1 & 1 \\
1 & -1
\end{psmallmatrix}$ and 
$h_c=(1/\sqrt{2}) \begin{psmallmatrix}
1 & 1 \\
i & -i
\end{psmallmatrix}$, we find $\rho_{\text{tot}}^{[1]} =(3/2)\mathbf{1}+(b_x+b_y+b_z)\sigma_x+b_z \sigma_y+(2b_x+b_z)\sigma_z$. Thus $\rho_{\text{tot}}^{[1]}= (3/2)\mathbf{1}$ exactly when $b_k=0$ for all $k=x, y, z$. Therefore a sufficient and necessary condition for the quantum Cram\'er-Rao  inequality to be saturated is
\begin{align} \label{eq:rho1mixed}
\rho_{z}^{[1]}=\frac{\mathbf{1}}{2}.
\end{align}
We can translate Eq.~\eqref{eq:rho1mixed} into an equivalent condition on photonic states. The most general pure $N$-particle two-mode bosonic state in the Fock basis is 
\begin{align} \label{eq:photonstate1}
\ket{\Psi}=\sum_{M=0}^N c_M\ket{M, N-M}.
\end{align}
Suppose the state in Eq.~\eqref{eq:photonstate1} is in the $HV$ basis. We can map it into its spin-state counterpart using Eq.~\eqref{eq:mapping}
\begin{align}
\ket{\Psi} \longleftrightarrow 
\ket{\xi_0}_{\text{spin}}=\sum_{M=0}^N c_M \frac{1}{\sqrt{\binom{N}{M}}}\sum_j \Pi_j \left(\ket{\uu}^{\otimes M} \otimes \ket{\dd}^{\otimes (N-M)}\right).
\end{align}
We identify $\rho_z^{[1]}$ with the one-particle reduced state of $\ket{\xi_0}_{\text{spin}}$. Again using Eq.~\eqref{eq:mapping}, we can map $\rho_z^{[1]}$ to its photonic counterpart $\rho^{[1]}$ and find
\begin{align} \label{eq:rho1spin}
&\rho^{[1]}=\frac{1}{N} \times [\nonumber \\
&\sum_{M=0}^N|c_M|^2M\ket{1,0}\bra{1,0}+\sum_{M=0}^N|c_M|^2(N-M)\ket{0,1}\bra{0,1}+ \nonumber \\
                &\sum_{M=0}^{N-1}c_M^*c_{M+1}\sqrt{(N-M)(M+1)}\ket{1,0}\bra{0,1}+ \nonumber \\
                 &\sum_{M=0}^{N-1}c_M c_{M+1}^*\sqrt{(N-M)(M+1)}\ket{0,1}\bra{1,0}].
\end{align}
Since $\bra{\Psi} a^{\dagger}a \ket{\Psi}=\sum_{M=0}^N|c_M|^2M=N-\bra{\Psi} b^{\dagger}b \ket{\Psi}$ and $\bra{\Psi} a b^{\dagger} \ket{\Psi}=\sum_{M=0}^{N-1}c_M^*c_{M+1}\sqrt{(N-M)(M+1)}$, Eq.~\eqref{eq:rho1spin} can be simplified to 
\begin{align}
\rho^{[1]}&=\frac{1}{N}(\bra{\Psi} a^{\dagger}a \ket{\Psi} \ket{1,0}\bra{1,0}+\bra{\Psi} b^{\dagger}b \ket{\Psi} \ket{0,1}\bra{0,1}) \nonumber \\
                   &+\frac{1}{N}(\bra{\Psi} a b^{\dagger}\ket{\Psi}\ket{1,0}\bra{0,1}+\bra{\Psi} a^{\dagger}b \ket{\Psi}\ket{0,1}\bra{1,0}).
\end{align}
This means the photonic equivalent to Eq.~\eqref{eq:rho1mixed} is
\begin{align}
\rho^{[1]}=\frac{1}{2} \ket{0,1}\bra{0,1}+\frac{1}{2}\ket{1,0}\bra{1,0}.
\end{align}
Thus the conditions for saturating the quantum Cram\'er-Rao  inequality (i.e. maximally-mixed one-particle density matrix) rewritten in the photonic form are
\begin{align} \label{eq:rho1conditions}
&\bra{\Psi} a^{\dagger} a \ket{\Psi} =\bra{\Psi} b^{\dagger} b \ket{\Psi}=\frac{N}{2}, \nonumber \\
&\bra{\Psi} a^{\dagger} b \ket{\Psi}=0=\bra{\Psi} a b^{\dagger} \ket{\Psi}.
\end{align}
The condition $\bra{\Psi} a^{\dagger} b \ket{\Psi}=0$ corresponds to an absence of first-order coherence for the state \cite{loudon}. An interpretation for an arbitrary state satisfying all the conditions of Eq.~\eqref{eq:rho1conditions} is that the intensities in each mode remain unaltered by any two-mode interferometer. 

The conditions in Eq.~\eqref{eq:rho1conditions} show that NOON states (except $N=1$) and Holland-Burnett states all saturate the quantum Cram\'er-Rao  inequality. In fact, all states of the form $\ket{M,N-M}+\ket{N-M,M}$ saturate the quantum Cram\'er-Rao inequality, \textit{except} when $M=(N \pm 1)/2$ (Yurke states for odd $N$ \cite{yurke1986input}). The Yurke states for even $N$ are $\ket{\Psi}=(1/\sqrt{2})(\ket{N/2, N/2}+\ket{N/2+1, N/2-1})$ \cite{yurke19862}, which also do not saturate the quantum Cram\'er-Rao inequality. 

The Fock states $\ket{N, 0}$ and $\ket{0, N}$ clearly violate the first equality in Eq.~\eqref{eq:rho1conditions} and thus also do not saturate the quantum Cram\'er-Rao inequality. Using Eq.~\eqref{eq:newQFIfull} for these states, we find $\tr(I^{-1})=3/(4N)$, which is shot-noise limited precision and agrees with known results in single-parameter estimation. 
\subsection{Optimal limit for the quantum Cram\'er-Rao inequality}
\label{sec:optimalproof}
To obtain the optimal precision in our protocol, we must find the lowest attainable value of $\tr(I^{-1})$. For this, we need first an argument from \cite{ballester2005} for the general form of $I$ for optimal states. We begin with the Cauchy-Schwarz inequality, which implies
$9=\tr(\mathbf{1})^2=[\tr(I^{-\frac{1}{2}}I^{\frac{1}{2}})]^2\leq \tr(I)\tr(I^{-1})$. The minimum value of $\tr(I^{-1})\vert_{\text{min}}$ occurs when the Cauchy-Schwarz inequality is saturated, or 
\begin{align} \label{eq:CSbound}
\tr(I^{-1})=\tr(I^{-1})\vert_{\text{min}}=\frac{9}{\tr(I)}.
\end{align}
This requires $I_{\alpha \beta} \propto \delta_{\alpha \beta}$. Next we note that the saturation condition $\rho_z^{[1]}=\mathbf{1}/2$ must also be satisfied for optimal $\rho_0$, and the general form of $I_{\alpha \beta}(\rho_0)$ from Eq.~\eqref{eq:newQFIfull} then becomes 
\begin{align} \label{eq:itotopt}
I_{\alpha \beta} (\rho_0)=2N[3 \delta_{\alpha \beta}+(N-1)\tr[\rho_{\text{tot}}^{[2]}(\sigma_{\alpha}\otimes \sigma_{\beta})]].
\end{align}
Inserting condition $I_{\alpha \beta} \propto \delta_{\alpha \beta}$ into Eq.~\eqref{eq:itotopt} yields 
\begin{align} \label{eq:rho2restriction}
\tr[\rho_{\text{tot}}^{[2]}(\sigma_{\alpha}\otimes \sigma_{\beta})] \propto \delta_{\alpha \beta}.
\end{align}
We identify the class of states $\rho_{\text{tot}}^{[2]}$ satisfying Eq.~\eqref{eq:rho2restriction} as follows. For the general form of $\rho_{\text{tot}}^{[2]}$ we write $\rho_{\text{tot}}^{[2]}=\rho^{[2]}_z+(h \otimes h) \rho^{[2]}_z (h \otimes h)+(h_c \otimes h_c) \rho^{[2]}_z (h^{\dagger}_c \otimes h^{\dagger}_c)$. We can expand $\rho_{z}^{[2]}$ in the basis of Pauli operators as $\rho_{z}^{[2]} =(\mathbf{1}\otimes\mathbf{1})/4+\sum_{j, k=x, y, z} c_{jk}(\sigma_j \otimes \sigma_k)$, where $c_{jk}=c_{kj}$ by the symmetry of $\rho^{[2]}_z$. We also note that the requirement for $\rho^{[1]}_z=\mathbf{1}/2$ eliminates contributions from terms in $\sigma_j \otimes \mathbf{1}$ and $\mathbf{1}\otimes \sigma_k$. 

The operators $h \otimes h$ and $h_c \otimes h_c$ act to permute off-diagonal terms of $\rho^{[2]}_z$ ($j \neq k$ or change sign for $c_{jk}$). From Eq.~\eqref{eq:rho2restriction} it can be readily verified that since the off-diagonal contributions to $\rho^{[2]}_{\text{tot}}$ must be zero, the off-diagonal contributions to $\rho^{[2]}_z$ must also disappear, i.e., $c_{jk}=0$ for $j \neq k$. From Eq.~\eqref{eq:rho2restriction} it also follows that $c_{xx}=c_{yy}$. Thus for the diagonal terms of $\rho^{[2]}_{\text{tot}}$ we find 
\begin{align} \label{eq:rhotot2K}
\rho_{\text{tot}}^{[2]}=(3/4)\mathbf{1}\otimes \mathbf{1}+K \sum_{k=x, y, z}(\sigma_k \otimes \sigma_k),
\end{align}
where $K=(2c_{xx}+c_{zz})$. Inserting Eq.~\eqref{eq:rhotot2K} into Eq.~\eqref{eq:itotopt}, we find $I_{\alpha \beta}=2N(3+4K(N-1))\delta_{\alpha \beta}$. From the optimality condition in Eq.~\eqref{eq:CSbound} this means optimal states should be those that maximize $K$. Since the eigenvalues of $\rho^{[2]}_{\text{tot}}$ are $3/4+K$ and $3/4-3K$, the maximum value of $K$ for maintaining physical states (i.e., eigenvalues range from $0$ to $3$) is $K=1/4$. Optimal states are therefore those satisfying
\begin{align}\label{eq:rhoz2new}
\rho_{z}^{[2]} 
=\frac{\mathbf{1}\otimes \mathbf{1}}{4}+\frac{1}{2}\left(\frac{1}{4}-c_{zz}\right)(\sigma_x \otimes \sigma_x+\sigma_y \otimes \sigma_y)+c_{zz}\sigma_z \otimes \sigma_z. 
\end{align}
For these optimal states, we have from Eq.~\eqref{eq:itotopt} that $I_{\alpha \beta}=2N(N+2)\delta_{\alpha \beta}$, which implies 
\begin{align} \label{eq:minI}
\tr(I_{\alpha \beta}^{-1})\vert_{\text{min}}=\frac{3}{2N(N+2)}.
\end{align}
We remark that this displays Heisenberg scaling $\tr(I_{\alpha \beta}^{-1}) \sim \mathcal{O}(N^{-2})$, signalling a quantum advantage in multi-parameter estimation. 

Following a similar derivation as in Sec.~\ref{sec:SaturateCR}, we can convert Eq.~\eqref{eq:rhoz2new} to its photonic counterpart using Eq.~\eqref{eq:mapping}
\begin{align} \label{eq:rho2zoptimal}
\rho^{[2]}=\left(\frac{1}{4}+c_{zz}\right)(\ket{2,0}\bra{2,0}+\ket{0,2}\bra{0,2})+2\left(\frac{1}{4}-c_{zz}\right)\ket{1,1}\bra{1,1}.
\end{align}
For the general photonic state $\ket{\Psi}$ in Eq.~\eqref{eq:photonstate1}, the equivalent photonic state to $\rho_z^{[2]}$ is
\begin{align}\label{eq:rho2photon2}
&\rho^{[2]}= \frac{1}{N(N-1)} \times [\nonumber \\
                  &\bra{\Psi} (a^{\dagger}a)(a^{\dagger}a-1)\ket{\Psi} \ket{2,0}\bra{2,0} \nonumber \\
&+\bra{\Psi} (b^{\dagger}b)(b^{\dagger}b-1)\ket{\Psi} \ket{0,2}\bra{0,2} \nonumber \\
                   &+2\bra{\Psi} (a^{\dagger}a)(b^{\dagger}b)\ket{\Psi} \ket{1,1}\bra{1,1} \nonumber \\
                   &+\bra{\Psi} a^{\dagger 2}b^2\ket{\Psi} \ket{0,2}\bra{2,0}+\bra{\Psi} (a^{\dagger 2}b^2)^{\dagger}\ket{\Psi} \ket{2,0}\bra{0,2} \nonumber \\
                   &+\sqrt{2}\bra{\Psi} (a^{\dagger}b)(b^{\dagger}b-1) \ket{\Psi} \ket{0,2}\bra{1,1} \nonumber \\
&+\sqrt{2}\bra{\Psi} ((a^{\dagger}b)(b^{\dagger}b-1))^{\dagger} \ket{\Psi} \ket{1,1} \bra{0,2} \nonumber \\
                  &+\sqrt{2}\bra{\Psi} (a^{\dagger}b)(a^{\dagger}a) \ket{\Psi} \ket{1,1}\bra{2,0} \nonumber \\
&+\sqrt{2}\bra{\Psi} ((a^{\dagger}b)(a^{\dagger}a))^{\dagger} \ket{\Psi} \ket{2,0}\bra{1,1}].
\end{align}
Comparing Eqs.~\eqref{eq:rho2zoptimal} and~\eqref{eq:rho2photon2}, the necessary and sufficient conditions for a state to be optimal (i.e., achieving the optimal quantum Cram\'er-Rao  inequality) can be written succinctly as
\begin{align} \label{eq:photonicconditions}
&\bra{\Psi} a^{\dagger}a \ket{\Psi}=\bra{\Psi} b^{\dagger} b\ket{\Psi}=\frac{N}{2}, \nonumber \\
&\bra{\Psi} a^{\dagger}a a^{\dagger} a\ket{\Psi}=\bra{\Psi} b^{\dagger}b b^{\dagger} b\ket{\Psi}, \nonumber \\
&\bra{\Psi} a^{\dagger} b \ket{\Psi}=0 =\bra{\Psi} a^{\dagger}a^{\dagger}a b \ket{\Psi}=\bra{\Psi} a^{\dagger}a^{\dagger}bb \ket{\Psi}=\bra{\Psi} a^{\dagger}b^{\dagger}bb\ket{\Psi}.
\end{align}
We observe that while the saturation of the Cram\'er-Rao  inequality depends only on the first-order correlations, the optimality conditions also depend on the second-order correlations $\bra{\Psi} a^{\dagger}a^{\dagger}a b \ket{\Psi}$, $\bra{\Psi} a^{\dagger}a^{\dagger}bb \ket{\Psi}$ and $\bra{\Psi} a^{\dagger}b^{\dagger}bb\ket{\Psi}$. This is related to the observation that in the corresponding spin system, the optimality condition is also a contraint on the two-particle reduced state, not only the one-particle reduced state. 

Using Eq.~\eqref{eq:photonicconditions}, it can easily be shown that optimal states include all Holland-Burnett states and states of the form $\ket{M,N-M}+\ket{N-M,M}$ (including NOON states) \textit{except} when $M=(N \pm1)/2$ (e.g. Yurke states, which do not saturate the quantum Cram\'er-Rao  inequality) \textit{and} when $M=N/2\pm1$ (e.g. NOON state when $N=2$ is suboptimal). 
\subsection{Comparison to alternative protocols}
\label{sec:comparisonschemes}
To compare the performance of our protocol with related proposals in the
literature, we now consider the general scenario in which an arbitrary probe
state $\rho _{0}$ with $N$ particles can be repeatedly prepared, and used
to obtain an estimate to an unknown unitary $U_{s}$. \ For this we adapt the
arguments in sections ~\ref{sec:SaturateCR} and ~\ref{sec:optimalproof} to compute the optimal precision
achievable using different choices for $\rho _{0}$, which includes the possibility of 
a collective measurement on all $N$
particles of the state $U_{S}^{\otimes N}\rho _{0}U_{S}^{\dag
\otimes N}$ for each run of the protocol. \ We note that that
the ability to prepare arbitrary $\rho _{0}$ here is comparable to the
ability in our protocol to prepare each of optimal $\rho _{x}$, $\rho _{y}$
and $\rho _{z}$  with $N$ particles ($\rho _{x}$, $\rho _{y}$ and $\rho _{z}$
are used for a third of the total measurements each). The spin versions of
optimal $\rho _{x}$, $\rho _{y}$ and $\rho _{z}$ we have discussed (e.g., spin
analogues of Holland-Burnett and NOON states under local rotations) are all
entangled. This is similar to the situation in single-parameter estimation
using entangled states of spins or qubits, for which it is well-known that
entanglement in the probe state plays a critical role in achieving
supersensitivity \cite{giovannetti2006quantum}.\\

To consider the case of arbitrary probe states for unitary estimation, we
must now modify our earlier definitions of the one and two-particle reduced
states to account for states $\rho _{0}$ which are not symmetric. \ We
define the `averaged' one-particle
reduced state of $\rho _{0}$ as $\tilde{\rho}_{0}^{\left[ 1\right]
}=(1/N)\sum_{i=1}^{N}\text{tr}_{\left[ i\right] }\left( \rho _{0}\right) $ and the
`averaged' two-particle reduced state of $%
\rho _{0}$ as $\tilde{\rho}_{0}^{\left[ 2\right] }=\left[ 1/\left( N\left(
N-1\right) \right) \right] \sum_{i>j}^{N}\left[ \text{tr}_{\left[ i,j\right]
}\left( \rho _{0}\right) +\text{tr}_{\left[ j,i\right] }\left( \rho _{0}\right) %
\right] $. \ Saturation of the quantum Cram\'{e}r-Rao bound in Eq.~\eqref{eq:taualphataubeta} then
reduces to $\text{tr}\left( \rho _{0}\left[ \tau _{\alpha },\tau _{\beta }\right]
\right) =\left( N/2\right) \text{tr}\left( \tilde{\rho}_{0}^{\left[ 1\right] }\left[
\sigma _{\alpha },\sigma _{\beta }\right] \right) =0$. \ By the same
argument in Section ~\ref{sec:SaturateCR}, this implies that $\tilde{\rho}_{0}^{\left[ 1\right]
}=\mathbf{1}/2$. \ Now the quantum Fisher information matrix for a general $N$-particle probe state satisfying this condition takes the form $%
I_{\alpha \beta }\left( \rho _{0}\right) =2N\left[ \delta _{\alpha \beta
}+\left( N-1\right) \right] \text{tr}\left[ \tilde{\rho}_{0}^{[2]}\left( \sigma
_{\alpha }\otimes \sigma _{\beta }\right) \right]$, using Eq.~\eqref{eq:totalqfisherinfo}. \
Optimality requires $I_{a\beta }\propto \delta _{\alpha \beta }$ and this
implies the constraint $\text{tr}\left[ \tilde{\rho} _{0}^{\left[ 2\right] }\left( \sigma
_{\alpha }\otimes \sigma _{\beta }\right) \right] \propto \delta _{a\beta }$%
, similar to Eq.~\eqref{eq:rho2restriction}. \ Since $\tilde{\rho}_{0}^{\left[ 1\right] }=\mathbf{1}/2$ and $%
\tilde{\rho}_{0}^{\left[ 2\right] }$ is symmetrised by definition, the
optimality constraint restricts 
\begin{align}\label{eq:33}
\tilde{\rho}_{0}^{\left[ 2\right] }=\frac{\mathbf{1}\otimes
\mathbf{1}}{4}+K\sum_{\alpha =x,y,z}\sigma _{\alpha }\otimes \sigma _{\alpha }.
\end{align}
Therefore $I_{a\beta }=2N\left[ 1+4K\left( N-1\right) \right] \delta
_{\alpha \beta }$ . \ Maximum precision comes from maximizing $K$ for
physical states (ie the eigenvalues for $\tilde{\rho}_{0}^{\left[ 2\right] }$
lie between 0 and 1). \ Since the eigenvalues of $\tilde{\rho}_{0}^{\left[ 2%
\right] }$ are $1/4+K$ and $1/4-3K$, the maximum is at $K=1/12$. \ Thus the
optimal precision is $\text{tr}\left( I_{\alpha \beta }^{-1}\right) |_{\text{min}%
}=9/\left[ 2N(N+2)\right] $. Note that if we used a separable state of $3N$ particles instead of a single $N$-particle state, this reduces precision to $\text{tr}\left( I_{\alpha \beta }^{-1}\right) |_{\text{min}%
}=3/\left[ 2N(N+2)\right]$. It is also clear from the dependence of $%
I_{\alpha \beta }\left( \rho _{0}\right) $ on the symmetrized quantities $%
\tilde{\rho}_{0}^{\left[ 1\right] }$ and $\tilde{\rho}_{0}^{\left[ 2\right] }
$ that there is no advantage to choices for $\rho _{0}$ which are not
symmetric. Since the symmetric state probes are those that can correspond to photonic 
states, this implies that spin protocols do not have an advantage over photonic protocols 
in terms of achieving optimal precision.\\

We can also compare one round of our protocol with the use of a single entangled probe state $\rho_0$ of $3N$ spin
particles. We can use the same argument in Section ~\ref{sec:SaturateCR} to derive the optimal precision in the latter case. Now the quantum Fisher information matrix takes the form $I_{\alpha \beta}(\rho_0)=2N[3 \delta_{\alpha \beta}+3(3N-1)\tr[\rho^{[2]}_0(\sigma_{\alpha}\otimes \sigma_{\beta})]$, using Eq.~\eqref{eq:totalqfisherinfo}, which contains extra contributions to the $\tr[\rho^{[2]}_0(\sigma_{\alpha}\otimes \sigma_{\beta})]$ term compared to Eq.~\eqref{eq:itotopt}. These come from correlations in $\rho_0$ that do not exist when $\rho_0=\rho_x \otimes \rho_y \otimes \rho_z$. The precision thus achievable would be $\text{tr}\left(
I_{\alpha \beta }^{-1}\right) |_{\text{min}}=3/\left[ 2N(3N+2)\right]$. 
This represents better precision but requires additional preparation
resources. \ On the other hand, when the restriction is made that probe
states can only be prepared with correlations on up to $N$ spin particles, the
optimal precision using our protocol versus repeated use of a single type of
probe state is the same, with $\text{tr}\left( I_{\alpha \beta }^{-1}\right) |_{%
\text{min}}=3/\left[ 2N(N+2)\right] $ when $3N$ spin particles are used. \ Note that
whatever protocol is used, the Fisher information analysis here can reliably
predict precision only when large number of measurements are used (roughly
100's of measurements would typically be used for each estimate of $U_{s}$). In this large number limit,
the precision scaling for both are similar.\\

Now we turn to optimal probe states, assuming that they can only be prepared
with correlations on up to $N$ spin particles. \ We have shown for our protocol that
any probe state whose two-particle reduced state in the $z$-basis obeys Eq.~\eqref{eq:rhoz2new} can achieve optimal precision. \ However, if an alternative protocol is
used with only one symmetric input state, as proposed in the recent protocol
in \cite{animesh}, the stricter Eq.~\eqref{eq:33} must be obeyed. \ An example of such a state
given in \cite{animesh} is $\sum_{i=x,y,z}\left( \left\vert +\sigma _{i}\right\rangle
^{\otimes N}+\left\vert -\sigma _{i}\right\rangle ^{\otimes N}\right) $
where $\left\vert \pm \sigma _{i}\right\rangle $ are the $\pm $eigenstates
of the Pauli matrics, in the large $N$ limit or when $N$ is a multiple of 8. For the analogous photonic protocol, these probe states are superpositions
of NOON states with respect to the polarisation bases HV, DA and RL
respectively.\\

Another way of satisfying Eq.~\eqref{eq:33} is given in \cite{ballester2005} and uses
protocols which allow for additional correlations between the probe state $%
\rho _{0}$ with $N$ spin particles and an ancilla state $\ket{i}_{\text{anc}}$ (which does not interact
with the unknown unitary). The optimal precision $\text{tr}\left( I_{\alpha \beta
}^{-1}\right) |_{\text{min}}=3/\left[ 2N(N+2)\right] $ is the same as above
with $N$-spin probe states. In \cite{ballester2005}, it is argued that all states proportional
to $\sum_{i=0}^{B-1}\left\vert s_{i}\right\rangle ^{\otimes N}\otimes
\left\vert i\right\rangle _{\text{anc}}$ satisfy Eq.~\eqref{eq:33} whenever the
states $\left\{ \left\vert s_{0}\right\rangle ,\cdots ,\left\vert
s_{B-1}\right\rangle \right\} $ for the probe satisfies $(1/B)\sum_{i=0}^{B-1}\ket{s_i}^{\otimes 2}\bra{s_i}^{\otimes 2}=(\mathbf{1}\otimes \mathbf{1})/4+(1/12)\sum_{\alpha=x,y,z}\sigma_{\alpha}\otimes \sigma_{\alpha}$ (also known as a 2-design) and the states 
$\left\{ \left\vert 0\right\rangle _{\text{anc}},\cdots ,\left\vert
B-1\right\rangle _{\text{anc}}\right\} $ for the ancilla are orthonormal. \
This is true for the Pauli basis where $\left\vert s_{0}\right\rangle
=\left\vert \sigma _{x}\right\rangle $, $\left\vert s_{1}\right\rangle
=\left\vert -\sigma _{x}\right\rangle $, $\left\vert s_{2}\right\rangle
=\left\vert \sigma _{y}\right\rangle $, $\left\vert s_{3}\right\rangle
=\left\vert -\sigma _{y}\right\rangle $, $\left\vert s_{4}\right\rangle
=\left\vert \sigma _{z}\right\rangle $ and $\left\vert s_{5}\right\rangle
=\left\vert -\sigma _{z}\right\rangle $. \ Another example is given by the
tetrahedral basis, such as $\left\vert s_{0}\right\rangle =\left\vert
0\right\rangle $, $\left\vert s_{1}\right\rangle =\left( 1/\sqrt{3}\right)
\left\vert 0\right\rangle -\left( 1/\sqrt{2}+i/\sqrt{6}\right) \left\vert
1\right\rangle $, $\left\vert s_{2}\right\rangle =\left( 1/\sqrt{3}\right)
\left\vert 0\right\rangle +\left( 1/\sqrt{2}-i/\sqrt{6}\right) \left\vert
1\right\rangle $, and $\left\vert s_{3}\right\rangle =\left( 1/\sqrt{3}%
\right) \left\vert 0\right\rangle +i\sqrt{2/3}\left\vert 1\right\rangle $. In addition, \cite{ballester2005} presents generalisations to higher-spin particles. Our
photon-spin mapping defined in Sec.~\ref{sec:photonmap} also generalises, and the results in \cite{ballester2005} for higher-spin particles can therefore be directly translated to
systems which use photonic states to probe linear-optical unitaries on 
$>2$ modes.\\

Finally we note that for any optimal choices for $%
\rho _{x}$, $\rho _{y}$ and $\rho _{z}$ in our protocol, which must satisfy
Eq.~\eqref{eq:rhoz2new}, the state $(1/3)\left( \rho _{x}\otimes \left\vert
0\right\rangle \left\langle 0\right\vert _{\text{anc}}+\rho _{y}\otimes
\left\vert 1\right\rangle \left\langle 1\right\vert _{\text{anc}}+\rho
_{z}\otimes \left\vert 2\right\rangle \left\langle 2\right\vert _{\text{anc}%
}\right) $ for the ancilla-assisted protocol automatically satisfies Eq.~\eqref{eq:33}
and achieves the same precision. \ In particular this shows that collective
measurements extending over $3N$ particles in our protocol (i.e., on $%
U_{S}^{\otimes N}\rho _{x}U_{S}^{\dag \otimes N}$, $U_{S}^{\otimes N}\rho
_{y}U_{S}^{\dag \otimes N}$ and $U_{S}^{\otimes N}\rho _{z}U_{S}^{\dag
\otimes N}$ together) cannot further improve the achievable precision.

\section{Linear-optical protocols using photon counting}
\label{sec:application}
Our previous analysis identifies the input states necessary for optimality. However, it does not indicate which projective measurements can be used to achieve this optimality. Here we focus on measurements that can be implemented in linear-optical experiments. In single-parameter estimation, Holland-Burnett and NOON states display Heisenberg scaling using photon-number-counting measurements. In this section, we find the precision achieved by general product states (which include Holland-Burnett states) and $N=2, 3$ NOON states using photon-number-counting measurements. 
\subsection{Fisher information for photon-number-counting measurements}
Consider the situation wherein the photonic state $\ket{\Psi}=\sum_{M=0}^N c_M \ket{M,N-M}$ in the Fock basis is used to probe an unknown linear-optical unitary U, with photon-number-counting measurements on each mode at the output. The probability of detecting the final state in the number state $\ket{M_d,N-M_d}$ is $|\bra{M_d, N-M_d}U\ket{\Psi}|^2$, where $M_d$ is an integer $0\leq M_d \leq N$. These probability distributions for $M_d=0, \cdots, N$ uniquely determine the Fisher information matrix.

Explicit forms for the probability distributions can be obtained using the Schwinger representation. This representation identifies $J_{x}=(a^{\dagger}b+a b^{\dagger})/2$, $J_y=(a^{\dagger}b-a b^{\dagger})/(2i)$ and $J_z=(a^{\dagger}a-b^{\dagger}b)/2$, where $J_{x}$, $J_y$, $J_z$ are the angular momentum operators acting on spin-$j$ states along the $x,y,z$ bases. The Schwinger representation therefore maps the photonic state $\ket{M, N-M}$ to a spin-$j$ state with quantum number $m$, represented by $\ket{j,m}$, where $j=N/2$ and $m=M-N/2$ \footnote{see \cite{sakurai} for a derivation}. For convenience, we write $U$ in terms of the Euler angle decomposition $U=\exp(i \psi_1 J_z)\exp(i \psi_2 J_y)\exp(i \psi_3 J_z)$, where $\psi_1$, $\psi_2$ and $\psi_3$ are Euler angles. Thus, given a known output state $\ket{j,m_d}$, where $m_d=M_d-N/2$, the probability distribution $|\bra{M_d, N-M_d}U\ket{\Psi}|^2$ can be expressed as a function of the three Euler angles. It becomes $P(m_d,\psi_1, \psi_2, \psi_3)=|\sum_{m=-j}^j c_m e^{i(\psi_3 m+\psi_1m_d)}d^j_{m_d,m}(\psi_2)|^2$. Here $d^j_{m_d,m}(\psi_2) \equiv \bra{j, m_d} \exp(i\psi_2 J_y) \ket{j,m}$ are the Wigner d-matrices (for a derivation of these see \cite{sakurai}, noting a different convention).

In our protocol, there are three types of input states and measurements corresponding to different polarisation bases $HV$, $DA$ and $RL$. The corresponding Fisher information matrices are denoted $F^{HV}$, $F^{DA}$ and $F^{RL}$ respectively. Due to the additivity condition in \cite{cramer1947mathematical}, the total Fisher information matrix is a sum of all three contributions $F^{HV}$, $F^{DA}$ and $F^{RL}$. 

We can relate $F^{DA}$ and $F^{RL}$ to $F^{HV}$ by observing how the probability distribution with respect to $DA$ (denoted $P_{DA}$) and $RL$ (denoted $P_{RL}$) can be transformed into the probability distribution with respect to $HV$ (denoted $P_{HV}$) by a change in Euler angles. Suppose $\{\psi_1, \psi_2, \psi_3\}$ are the Euler angles corresponding to the transformation with respect to the $HV$ basis. Then we can write $P_{DA}(m_d, \psi_1, \psi_2, \psi_3) \equiv |\bra{j, m_d}_{DA}U\ket{\Psi}_{DA}|^2=|\bra{j, m_d}_{HV}U'\ket{\Psi}_{HV}|^2 \equiv P_{HV}(m_d, \psi'_1, \psi'_2, \psi'_3)$ where $U'=\exp(i \psi'_1 J_z)\exp(i \psi'_2 J_y)\exp(i \psi'_3 J_z)$. Thus the Euler angles  $\{\psi'_1, \psi'_2, \psi'_3\}$ are defined by $P_{HV}(m_d,\psi'_1, \psi'_2, \psi'_3)=P_{DA}(m_d,\psi_1, \psi_2, \psi_3)$. Similarly, the Euler angles $\{\psi''_1, \psi''_2, \psi''_3\}$ are defined by $P_{HV}(m_d,\psi''_1, \psi''_2, \psi''_3)=P_{RL}(m_d,\psi_1, \psi_2, \psi_3)$. 

We define matrices $W'$ and $W''$ by $W'_{\alpha \beta} \equiv \partial \psi'_{\alpha}/\partial \psi_{\beta}$ and $W''_{\alpha \beta} \equiv \partial \psi''_{\alpha}/\partial \psi_{\beta}$. Therefore 
\begin{align}\label{eq:ftransform}
&F^{DA} (\psi_1, \psi_2, \psi_3)=W'^T F^{HV}(\psi'_1, \psi'_2, \psi'_3) W' \nonumber \\
&F^{RL} (\psi_1, \psi_2, \psi_3)=W''^T F^{HV}(\psi''_1, \psi''_2, \psi''_3) W''. 
\end{align}

For practical purposes including computing probability distributions and performing estimation from data, it is convenient to use the Euler angle parameters (or any other simple parametrization).  However, we have seen that locally-independent
parameters are better suited to quantifying precision.

We can transform the Fisher information matrix in the Euler parameterisation $F_{Euler}$ into the locally-independent parameterisation (with parameters $\{\theta_1, \theta_2, \theta_3\}$) by $F=J^T F_{Euler} J$, where matrix $J$ is defined as $J_{\alpha \beta}\equiv \partial \psi_{\alpha}/\partial \theta_{\beta}$. Since the locally-independent parameters are defined by $t_{\alpha} \equiv\sigma_m/\sqrt{2}$, we can find $J$ using
\begin{equation} \label{eq:sigmaj1}
\frac{\sigma_{\alpha}}{\sqrt{2}}= t_{\alpha}=iU_s^{\dagger}\sum_k \frac{\partial U_s}{\partial \psi_k} J_{k \alpha}.
\end{equation}
The last term in Eq.~\eqref{eq:sigmaj1} can be computed by replacing $U_s$ with its matrix representation $\mathcal{M}$ (see Appendix~\ref{app:photonmap}), expressed in its Euler angle decomposition as $\mathcal{M}=\exp(i \psi_1 \sigma_z/2)\exp(i \psi_2\sigma_y/2)\exp(i\psi_3\sigma_z/2)$. Since the optimal precision is characterised by $\tr(F^{-1})$, we find $\tr(F^{-1})=\tr(V F^{-1}_{Euler})$, where 
\begin{align}
V=(J^{-1})^T J^{-1}=\frac{1}{2}
\begin{pmatrix}
1 & 0 & \cos(\psi_2) \\
0 & 1 & 0 \\
\cos(\psi_2) & 0 & 1
\end{pmatrix}.
\end{align}
$V$ becomes a simple rescaling factor around $\psi_2=\pm \pi/2$. Thus we expect the Euler parameters to behave similarly to locally-independent parameters around $\{\psi_1, \psi_2=\pm \pi/2,\psi_3\}$. Suppose we denote infinitesimal deviation from these Euler angles by $\{\delta \psi_1, \delta \psi_2, \delta \psi_3\}$. Then $U_s (\psi_1+\delta \psi_1,\pm \pi/2+\delta \psi_2, \psi_3+\delta \psi_3) \simeq U_s (\psi_1, \psi_2=\pm \pi/2, \psi_3) R_z^{\dagger}(\mathbf{1}\pm i \delta \psi_1 \sigma_x/2+i \delta \psi_2 \sigma_y/2+i \delta \psi_3 \sigma_z/2)R_z$ where $R_z=\exp(i\psi_3 \sigma_z/2)$ is a rotation acting on the generators. Since the sign changes and fixed rotation $R_z$ preserve the essential features of the locally-independent parameterisation, this confirms that Euler parameterisation is equivalent to locally-independent parameterisation up to a scaling factor.

We also note that $\det(V)=\sin(\psi_2)^2/8$, which goes to zero at points $\psi_2=0$ and $\psi_2=\pm \pi$. Thus, at these points, the inverse Fisher information with respect to the Euler parameterisation diverges. This can happen when one cannot gather any information about any one (or more) parameter(s). For instance, when $\psi_2=0$, $U_s=\exp(i(\psi_1+\psi_3) \sigma_z/2)$. Here one is estimating only two parameters $\psi_1+\psi_3$ and $\psi_2=0$ instead of three parameters. Similarly when $\psi=\pm \pi$, one is estimating two parameters $\psi_3-\psi_1$ and $\psi_2=\pm \pi$ instead of three parameters, since $U_s=\exp(\pm i (\pi/2) \sigma_y) \exp(i(\psi_3-\psi_1) \sigma_z/2)$. However, this singularity is an artificial product of the Euler parameterisation and not essential to the estimation protocol. Other parameterisations will lead to singularities at different points. To distinguish these artificial singularities from genuine singularities that might arise from actual limitations of the estimation scheme, we can use alternative choices of parameterisation.

The tools in this section are sufficient to find the optimal precision possible in our protocol when using photon-number-counting measurements and $N$-photon probe states. 
\subsection{Product states}
We now examine the precision achievable by product states $\ket{M,N-M}$ using photon-number-counting meausurements. In single-parameter estimation, the Fisher information is $F_{single}=N+2M(N-M)$, corresponding to $U_s=\exp(i \psi_2 J_y)$ for fixed $\psi_1=0=\psi_3$ \cite{durkin2007}. This is maximised for Holland-Burnett states (i.e., $M=N/2$), where $F_{single}=N(N+2)/2$. For multiparameter estimation using product states, the Fisher information matrix $F^{HV}(\psi_1, \psi_2, \psi_3)$ has only one non-zero element, $F^{HV}_{22}(\psi_2)$, since the probability distributions using product states depend on $\psi_2$, but not $\psi_1$ or $\psi_3$.  Hence, the scaling of $\tr(F^{-1})$ can be shown, using Eq.~\eqref{eq:ftransform}, to be equivalent to that of single-parameter estimation, where
\begin{align}\label{eq:newtrvw22}
\tr(F^{-1})=\frac{1}{F_{single}}\tr(V \tilde{W}^{-1}).
\end{align}
Here the matrix $\tilde{W}$ is defined by $\tilde{W}_{\alpha \beta}=\delta_{2\alpha} \delta_{2 \beta}+W'_{2\alpha}W'_{2\beta}+W''_{2\alpha } W''_{2 \beta}$. It captures the effect of the three sets of polarisation bases used and is independent of the input state. The matrix $\tilde{W}$ can be easily computed using $\cos^2(\psi'_2/2)=\cos^2(\psi_2/2)\cos^2((\psi_1+\psi_3)/2)+\sin^2(\psi_2/2)\sin^2((\psi_1-\psi_3)/2)$ and $\cos^2(\psi''_2/2)=\cos^2(\psi_2/2)\cos^2((\psi_1+\psi_3)/2)+\sin^2(\psi_2/2)\cos^2((\psi_1-\psi_3)/2)$. 

Our numerical simulations show the minimum value of $\tr(V \tilde{W}^{-1})$ to be $3/2$. This indicates that Holland-Burnett states are the only product states under photon-number-counting measurements that achieve an identical scaling in $N$ with the theoretical optimal in Eq.~\eqref{eq:minI}, with the minimum $\tr(F^{-1})=3/(N(N+2))$. However, this is a factor of two higher than the predicted value for optimal measurements. It is interesting to compare to single-parameter estimation, where the optimal precision for Holland-Burnett states do not in fact share exactly the same scaling in $N$ as the theoretical optimal scaling of $1/N^2$. 

Holland-Burnett states are thus strong candidates for practical implementation of our protocol. As well as providing near-optimal precision using photon-number-counting measurements, they are more experimentally accessible than NOON states for $N>2$ \cite{datta2011quantum} and also perform better in the presence of photon loss. 

We also observe that the precision in multi-parameter estimation $\tr(F^{-1})$ is \textit{dependent} on the parameters of the unknown unitary. This is different to single-parameter estimation for Holland-Burnett states, where the Fisher information is independent of the unknown parameter. We can see this dependence by first noting the positions of the minima, which occur at $\{\psi_1, \psi_2, \psi_3\}=\{0, \pm \pi/2, \pm \pi/2\}$, $\{\pm \pi/2, \pm \pi/2, 0\}$, $\{\pm \pi, \pm \pi/2, \pi/2\}$, $\{\pm 3\pi/2, \pm \pi/2, 0\}$. These minima occur at $\psi_2=\pm \pi/2$, where the Euclidean parameterisation coincides with the locally-independent parameterisation. If we take $U_s=\begin{psmallmatrix}
a+ib & c+id \\
-c+id & a-ib
\end{psmallmatrix}$, where $a,b,c,d \in \Re$ and $a^2+b^2+c^2+d^2=1$, then the minima occur when $|a|=|b|=|c|=|d|=1/2$. This coincides with the results found in \cite{hugotomo} that use process fidelity instead of Fisher information. \\

\begin{figure}[ht!]
\centering
\includegraphics[scale=0.3]{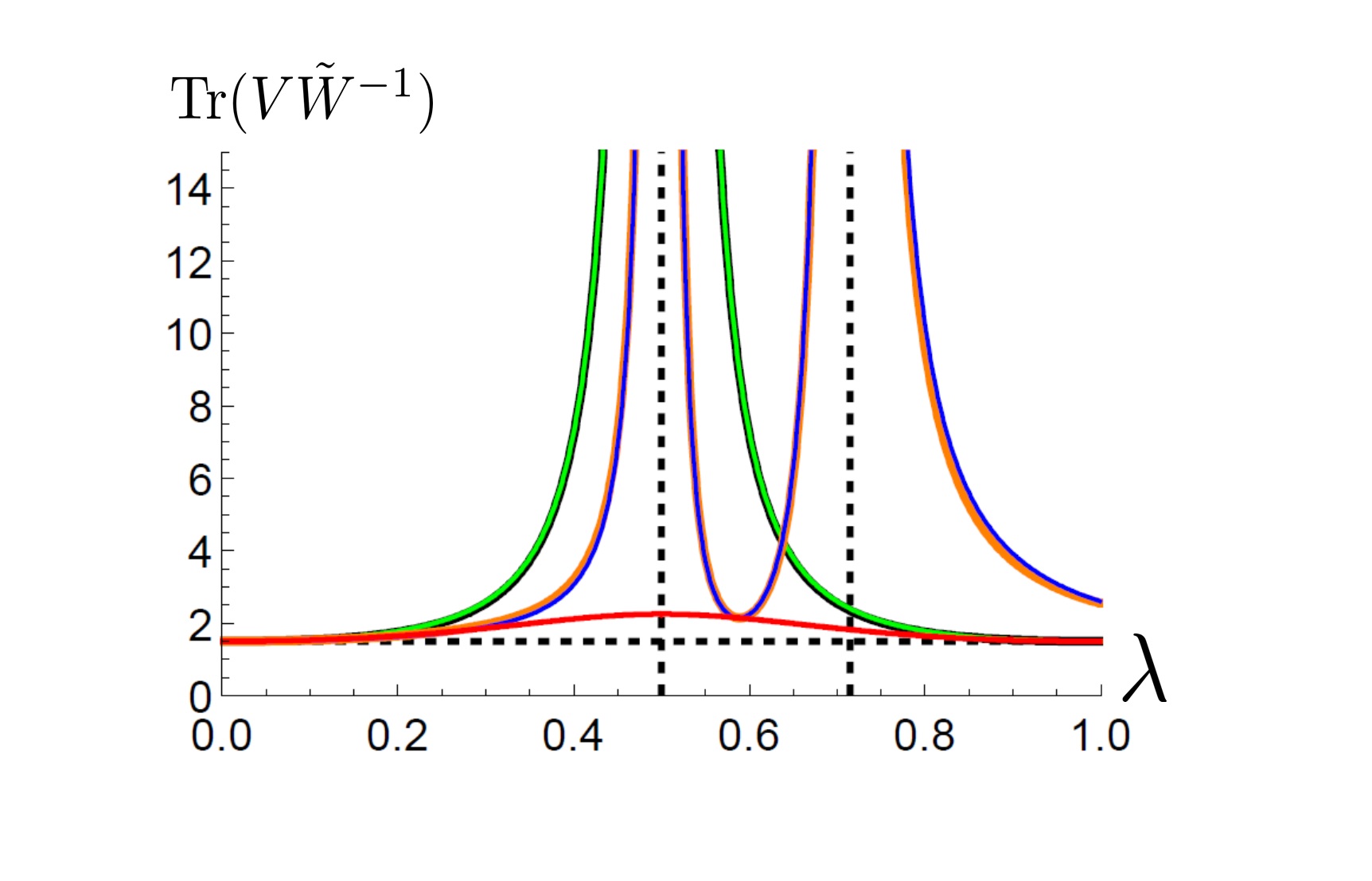}
\caption[\textit{Behaviour of $\tr(V \tilde{W}^{-1})$ for different paths $P_i (\lambda)$ along $\lambda$}.]{\label{fig:trvw}Behaviour of $\tr(V \tilde{W}^{-1})$ for five different families of unitaries along paths $P_i (\lambda)$ (where $i=1, 2, 3, 4, 5$)
with respect to $\lambda$, in the $\{a,b,c,d\}$ parameter space for $U_s=\begin{psmallmatrix} a+ib & c+id \\
-c+id &a-ib \end{psmallmatrix}$. The paths $P_1 (\lambda)$ (black) and $P_2(\lambda)$ (green) give near identical $\tr(V \tilde{W}^{-1})$ with a divergence at $\lambda=0.5$. The paths $P_3 (\lambda)$ (orange) and $P_4 (\lambda)$ (blue) also give near identical $\tr(V \tilde{W}^{-1})$ with two divergences, at $\lambda=0.5$ and $\lambda=0.5/0.7$ (vertical dashed black). The path $P_5 (\lambda)$ (red) does not pass through any divergences and always remains close to the minimum value of $\tr(V \tilde{W}^{-1})=3/2$ (horizontal dashed black). See text for more details.}
\end{figure}

We can examine the dependence of $\tr(F^{-1})$ on $U_s=\begin{psmallmatrix} a+ib & c+id \\
-c+id &a-ib \end{psmallmatrix}$ by showing how $\tr(V \tilde{W}^{-1})$ changes with respect to unitaries that are near the minimal unitary $u_{min}=\{a,b,c,d\}=\{1/2,1/2,1/2,1/2\}$ (i.e., the unitary where $\tr(F^{-1})$ attains its lowest value). Unitaries near $u_{min}$ can be accessed from $u_{min}$ through different paths in the $\{a,b,c,d\}$ parameter space. Each point along a path represents a different unitary we want to estimate and we can study the behaviour of $\tr(V \tilde{W}^{-1})$ at each unitary along each path. We can define such five paths $P_i (\lambda) \equiv p_i(\lambda)/||p_i(\lambda)||$ where $i=1, 2, 3, 4, 5$ and each path is parameterised by $\lambda \in [0,1]$ in the following way: $p_1=u_{min}-\{\lambda, 0, 0, 0\}$, $p_2=u_{min}-\{\lambda, 0, \lambda, 0\}$, $p_3=u_{min}-\{\lambda, 0.7 \lambda, \lambda, 0\}$, $p_4=u_{min}-\{0.7 \lambda, \lambda, 0.7 \lambda, 0\}$ and $p_5=u_{min}-\{\lambda, \lambda, \lambda, 0\}$. See Fig.~\ref{fig:trvw} for a plot showing how $\tr(V \tilde{W}^{-1})$ changes along $p_i(\lambda)$. There are two divergences that occur at $\lambda=0.5$ and $\lambda=0.5/0.7$, which correspond to at least one of $a, b, c, d$ having value zero. 
 
Dependence of the Fisher information on the unitary matrix itself is in fact common in single-parameter estimation when not dealing with so-called path-symmetric states (which include Holland-Burnett and NOON states \cite{hofmann2009}) or when the effects of experimental imperfections are accounted for. Adaptive schemes such as in \cite{adaptive1, adaptive2} can be used to optimize precision given these dependencies. 
\subsection{NOON states}
We now examine the precision achieved by $N=2, 3$ NOON states under photon-number-counting measurements. We restrict our attention to small-$N$ NOON states since proposed schemes for efficient generation of large-$N$ NOON states achieve fidelity considerably less than 1 \cite{hofmann2007high} \footnote{We note the fidelities achieved in the scheme in \cite{hofmann2007high} are more than $90\%$ for NOON components with high photon number.}, or require feedforward \cite{cable2007efficient, cable2011formation} which is technically-challenging. Although all large-$N$ quantum states with high phase sensitivity are very sensitive to photon losses, the problem is particularly acute for NOON states, for which the loss of a single photon due to losses acting independently on both modes causes a complete loss of phase sensitivity.

\begin{figure}[ht!]
\centering
\includegraphics[scale=0.3]{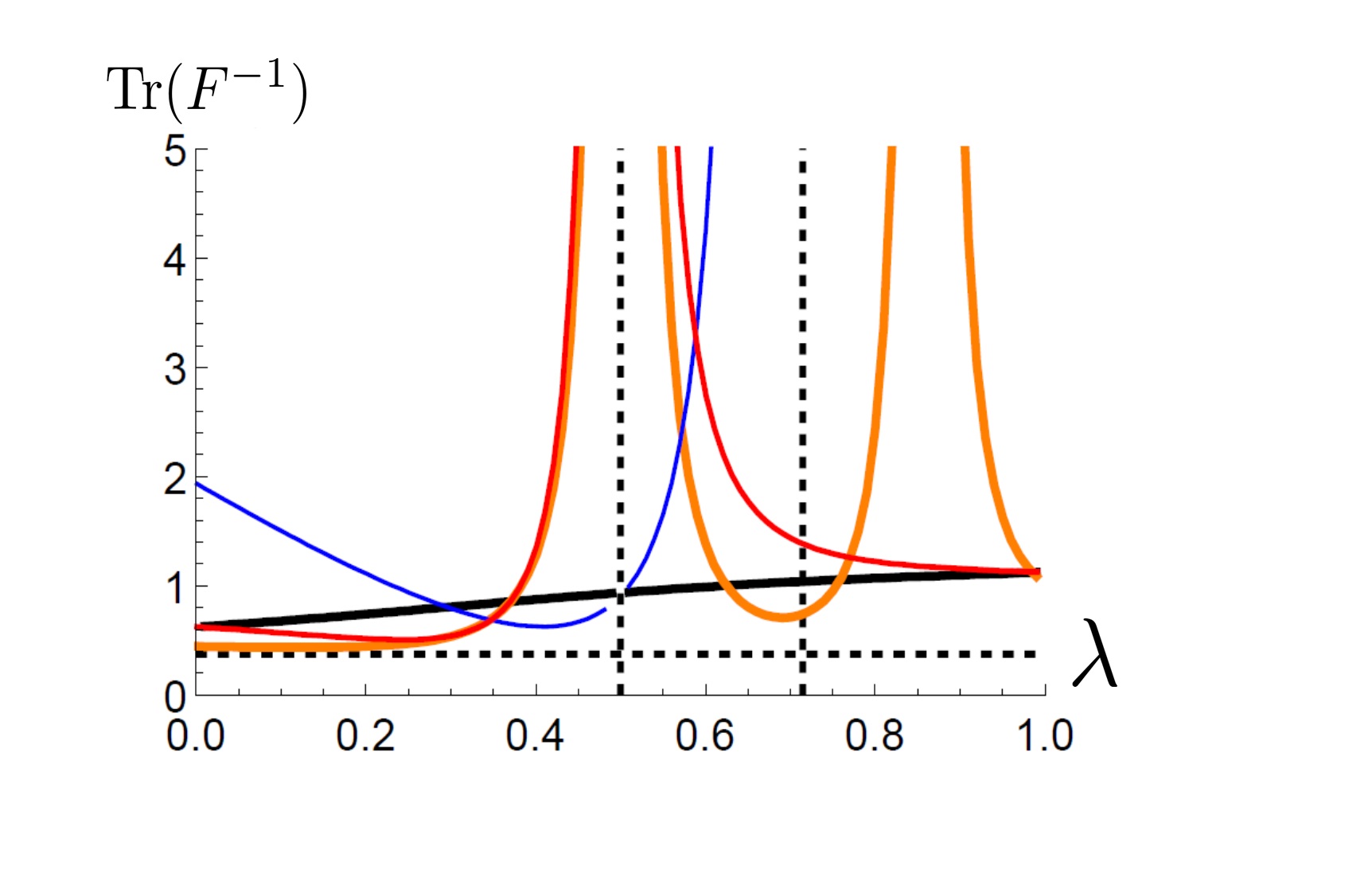}
\caption[\textit{Behaviour of $\tr(F^{-1})$ for different paths $P_i (\lambda)$ along $\lambda$ for $N=2$ NOON state}.]{\label{fig:noon2}Behaviour of $\tr(F^{-1})$ for four different families of unitaries along paths $P_i (\lambda)$ (where $i=1, 3, 4, 5$)
with respect to $\lambda$, in the $\{a,b,c,d\}$ parameter space for $U_s=\begin{psmallmatrix} a+ib & c+id \\
-c+id &a-ib \end{psmallmatrix}$, for $N=2$ NOON state. $\text{tr}(F^{-1})$ is given for trajectories $p_1(\lambda)$ (black), $p_3(\lambda)$ (orange), $p_4(\lambda)$ (blue) and $p_5(\lambda)$ (red).  $\text{tr}(F^{-1})$ is ill-conditioned along $p_2(\lambda)$, and so this trajectory is not shown.  A lower bound for precision for all the trajectories is given by $\text{tr}(F^{-1})=0.375$, which is shared by the $N=2$ Holland-Burnett state (horizontal dashed black).  Points where at least
one of $a$, $b$ ,$c$, $d$ have value zero are indicated at $\lambda=0.5$ and $\lambda=0.5/0.7$ (vertical dashed black), revealing differences in the locations of some divergent behavior compared to the $N=2$ Holland-Burnett state in Fig.~\ref{fig:trvw}.}
\end{figure}

\begin{figure}[ht!]
\centering
\includegraphics[scale=0.3]{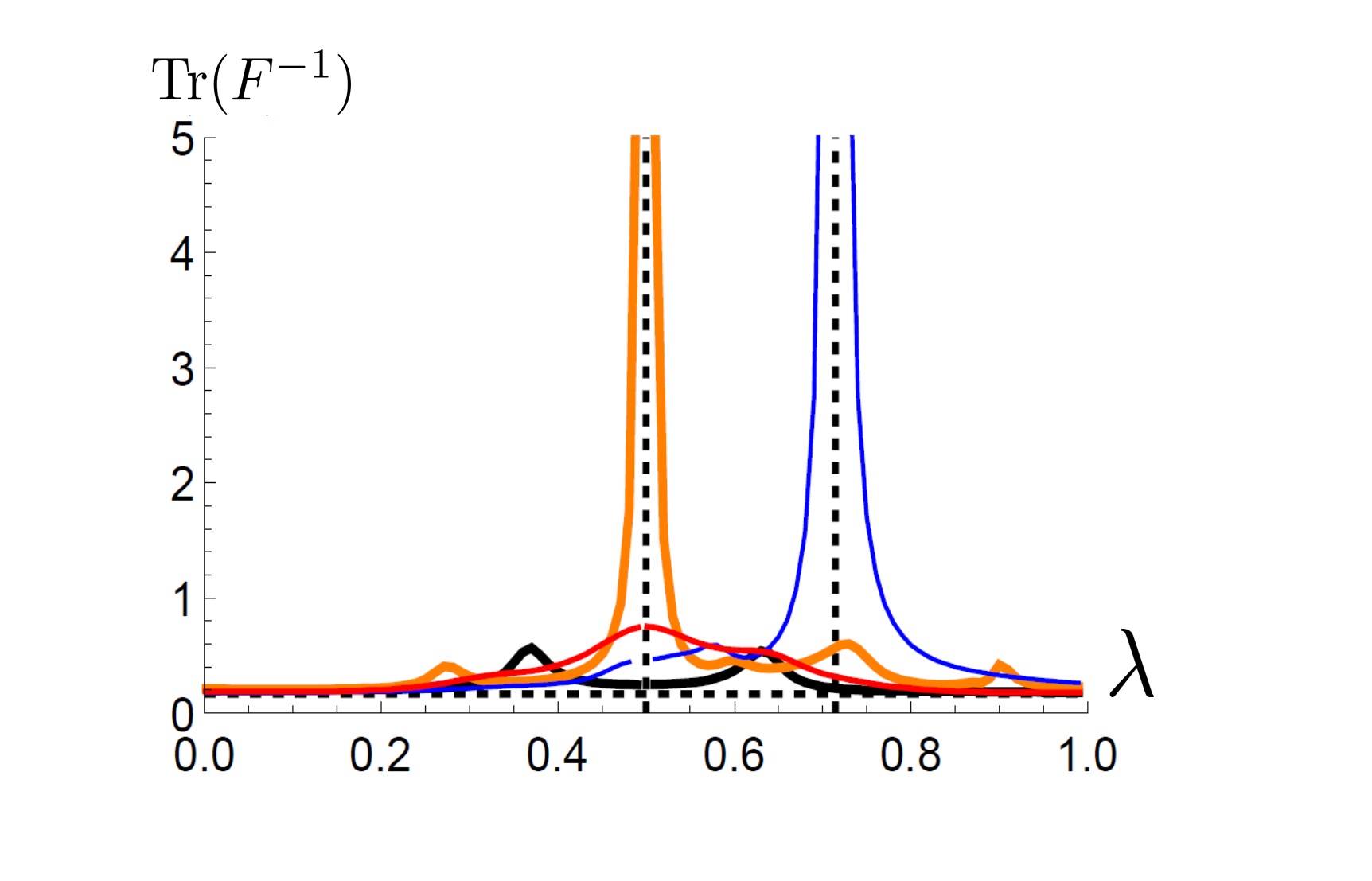}
\caption[\textit{Behaviour of $\tr(F^{-1})$ for different paths $P_i (\lambda)$ along $\lambda$ for $N=3$ NOON state}.]{\label{fig:noon3}Behaviour of $\tr(F^{-1})$ for four different families of unitaries along paths $P_i (\lambda)$ (where $i=1, 3, 4, 5$)
with respect to $\lambda$, in the $\{a,b,c,d\}$ parameter space for $U_s=\begin{psmallmatrix} a+ib & c+id \\
-c+id &a-ib \end{psmallmatrix}$, for $N=3$ NOON state. The paths $P_1 (\lambda)$ (black) and $P_5 (\lambda)$ (red) stay near $\tr(F^{-1})=0.167$ (horizontal dashed black). The path $P_3 (\lambda)$ (orange) stays close to $\tr(F^{-1})=0.167$ except near a divergence at $\lambda=0.5$ (vertical dashed black). The path $P_4 (\lambda)$ (blue) exhibits a divergence at $\lambda=0.5/0.7$ (vertical dashed black). $\tr(F^{-1})$ is ill-conditioned along the whole path $P_2(\lambda)$ and is not pictured.}
\end{figure}

We find $\tr(F^{-1})$ for $N=2$ and $N=3$ NOON states by directly computing $F^{HV}$, $F^{DA}$ and $F^{RL}$ from the corresponding probability distributions. Plots for $\tr(F^{-1})$ along the paths $P_i (\lambda)$ for the $N=2$ NOON state are given in Fig.~\ref{fig:noon2} and for the $N=3$ NOON state in Fig.~\ref{fig:noon3}. 

For the case of $N=2$, we note that although a unitary $U_s=h_c$ can be used to convert the Holland-Burnett state into the NOON state, this transformation does not commute with $h$, used for the $DA$ polarisation-basis measurement.  Hence, the general behaviour of the precision computed for the two states should be expected to be quite different. 

Numerical investigation based on a random search over 1000 Haar-random matrices (that represent unitary transformations we want to estimate) yields a minimum $\tr(F^{-1})=0.377$ for $N=2$ NOON states. This is very close to the $N=2$ Holland-Burnett state with $\tr(F^{-1})=0.375$. Neither of these states are optimal which requires $\tr(F^{-1})=0.1875$. 

A similar search for $N=3$ NOON states yields a minimum $\tr(F^{-1})=0.167$. This is slightly better precision than the value $0.2$, which is given for $N=3$ by the formula $3/N(N+2)$ (which applies for Holland-Burnett states with even $N$). Note that the optimal value $\tr(F^{-1})=0.1$ is not achieved.

Thus the optimality of $N=2$ and $N=3$ NOON states together with photon-number-counting measurements for single-parameter estimation no longer holds true in the multi-parameter estimation protocol that we consider. Furthermore, the best achievable precision by these NOON states is similar to that achievable using Holland-Burnett states. 
\subsection{$N=2,3$ states and photon-number-counting measurements}
It is interesting to consider if there exist $N=2,3$ states which can reach the optimal bound on precision in Eq.~\eqref{eq:minI} using photon-number-counting measurements. For $N=2$, the Holland-Burnett state is the only state whose precision saturates the optimal bound in Eq.~\eqref{eq:minI} for general measurements and for $N=3$, the NOON state is the only such optimal state (see Appendix~\ref{app:noon3optimal}). Due to the results in 4.2 and 4.3, this means that there is no $N=2,3$ state that can saturate the optimal precision using photon-number-counting measurements.\\

However, there may still exist states that can perform better than Holland-Burnett and NOON states under photon-number-counting measurements only. It is interesting to explore the precision attained by these states, but finding these states and simulating their precision is difficult. It requires the computation of 18 independent terms for the Fisher information matrix (each of which is dependent on multi-photon statistics), that correspond to three independent measurements. From the examples of Holland-Burnett and NOON states, we have seen that there is also a high sensitivity of the precision on the unitary and the task for optimising over random unitaries is already complicated even for any given $N=2,3$ state. Furthermore, there are no known methods for generating arbitrary states with fixed total photon number in linear-optical measurements (especially for $N>2$). If a $N=3$ state with better precision under photon-number-counting measurements were found, it may be difficult to experimentally generate. This is similar to the case in single-parameter estimation, where in general there are only limited cases when an experimentally-motivated measurement does saturate the quantum Cram\'er-Rao bound.\\

We also note that, for arbitrary $N$, the precision $\text{tr}(F^{-1})$ for Holland-Burnett states under photon-number-counting measurements only differs from the optimal precision by a factor of $2$. Therefore, practical considerations may make Holland-Burnett states more desirable to use over other possible states, which in principle provide marginally better performance but which would be more difficult to generate. While Holland-Burnett and NOON states are not optimal under photon-number-counting measurements, it is an interesting question for future investigation what kinds of measurements are optimal for these states. However, it is unclear if there are any `simple' optimal measurements, as there are few experimentally available options other than photon-number-counting measurements.

\section{Summary and outlook}
\label{sec:summary}
In summary, we have developed a formalism to study quantum-enhanced $SU(2)$-estimation using $N$-particle photonic states. We have derived easy-to-use, necessary and sufficient conditions that these photonic states must satisfy to achieve the optimal precision in $SU(2)$-estimation. We also interpreted these results in terms of photon interferometry. Our results showed some key differences between multi-parameter and single-parameter estimation. 

We found that, unlike single-parameter estimation (without loss and decoherence), Holland-Burnett states and NOON states provide a strongly unitary-dependent precision, making adaptive measurements essential. Holland-Burnett states are optimal, although they are only near optimal using photon-number-counting measurements. This makes Holland-Burnett states experimentally preferable to high-$N$ NOON states which are difficult to generate. In addition, $N=2$ and $N=3$ NOON states do not achieve optimal precision under photon-number-counting measurements. 

Our results clearly show how multi-parameter estimation is not a simple generalisation of single-parameter estimation and thus invites further theoretical study. 

By using a mapping between photonic and spin systems, we have argued how to relate multi-parameter estimation schemes for photonic and spin systems. As a first application, we have shown how the spin analogue to our photonic protocol allows more optimal states than an earlier proposal for spin systems. The mapping we have described will potentially inform improvements for future protocols for both systems.

Extending our results to more practically important schemes presents some exciting challenges. For example, the use of multi-mode squeezed states and homodyne measurements could potentially enable quantum enhancement for unitary estimation at high-$N$. Extensions of the general approach in this paper to $SU(d)$ estimation for $d>2$ may have implications for quantum computing models like boson sampling. An important question for future study is how the potential for quantum-enhanced precision changes for non-unitary processes that include the effects of photon loss and decoherence. This question has been a focal point for work on single-parameter estimation in recent years. In particular, the Heisenberg limit has been largely superseded by revised scaling laws that account for imperfections and which reveal much-reduced potential for supersensitivity using non-classical probe states \cite{escher2011general}. 

Furthermore, there are many alternative notions of quantum enhancement compared to the one used in this paper, that should be considered for different applications with specific restrictions on physical resources.  In particular, recent work on single-parameter estimation shows the importance of critically comparing the performance of single and multi-pass protocols using non-classical and classical probe states where there is a requirement to maximize precision per absorbed photon, as is key when measuring fragile samples \cite{birchall2016quantum}.
\ack
N. L. thanks Benjamin Yadin and Mihai Vidrighin for interesting discussions. N.L. also acknowledges support by the Clarendon Fund and Merton College of the University of Oxford. This material is also based on research funded by the Singapore National Research Foundation under NRF Award NRF-NRFF2013-01. H.C. thanks X.-Q. Zhou and M. Ballester for discussions. H.C. acknowledges support from the Engineering and Physical Sciences Research Council (EPSRC, UK) and US Army Research Office (ARO) Grant W911NF-14-1-0133. No data were created during this study.  


\begin{appendices}
\section{Photonic to spin system mapping}
\label{app:photonmap}
Using the mapping in Eq.~\eqref{eq:mapping} we can describe a transformation of the two-mode photonic state under unitary operator $U$ in terms of the evolution of $N$ spin-$1/2$ particles. Let each spin-$1/2$ particle transform under the $2 \times 2$ unitary matrix represented by $\mathcal{M}$. Then $\mathcal{M}$ and $U$ are related by 
\begin{align} \label{eq:spinphoton} \tag{A.1}
\begin{pmatrix}
\tilde{a}^{\dagger} \\
\tilde{b}^{\dagger}
\end{pmatrix} \equiv 
\begin{pmatrix}
U a^{\dagger} U^{\dagger} \\
U b^{\dagger} U^{\dagger}
\end{pmatrix}=
\begin{pmatrix}
\alpha & \beta \\
\gamma & \delta
\end{pmatrix} 
\begin{pmatrix}
a^{\dagger} \\
b^{\dagger}
\end{pmatrix} \equiv
\mathcal{M}^{T} \begin{pmatrix}
a^{\dagger} \\
b^{\dagger}
\end{pmatrix},
\end{align}
where $U$ is confined to a linear-optical process. Here $\mathcal{M}$ can be interpreted as acting on a single-photon state. 

We begin with the correspondence between the creation operators of the photonic and spin states $a^{\dagger} \leftrightarrow a^{\dagger}_{\uu}$, $b^{\dagger} \leftrightarrow a^{\dagger}_{\dd}$ from our mapping. Thus, after unitary evolution, we have the correspondence $Ua^{\dagger}U^{\dagger}=\tilde{a}^{\dagger} \leftrightarrow U_s a^{\dagger}_{\uu} U_s^{\dagger} \equiv \tilde{a}^{\dagger}_{\uu}$ and $Ub^{\dagger}U^{\dagger}=\tilde{b}^{\dagger} \leftrightarrow U_s a^{\dagger}_{\dd} U_s^{\dagger} \equiv \tilde{a}^{\dagger}_{\dd}$, where $U_s$ is the unitary operator acting on the spin degrees of freedom. To find the correspondence between $U_s$ and $U$, we can see from Eq.~\eqref{eq:spinphoton} that $U\ket{1,0}=U a^{\dagger} U^{\dagger} \ket{0,0}=\alpha a^{\dagger} \ket{0,0}+\beta b^{\dagger} \ket{0,0} \leftrightarrow \alpha \ket{\uu}+\beta \ket{\dd}=U_s a^{\dagger}_{\uu} U_s^{\dagger} \ket{0}=U_s a^{\dagger}_{\uu} \ket{0}=U_s \ket{\uu}$ and $U\ket{0,1}=U b^{\dagger} U^{\dagger} \ket{0,0}=\gamma a^{\dagger} \ket{0,0}+\delta b^{\dagger} \ket{0,0}\leftrightarrow \gamma \ket{\uu}+\delta \ket{\dd}=U_s \ket{\dd}$. Therefore we can write 
\begin{align} \tag{A.2}
\begin{pmatrix}
\tilde{a}^{\dagger}\\
\tilde{b}^{\dagger}
\end{pmatrix}\equiv
\begin{pmatrix}
U a^{\dagger} U^{\dagger} \\
U b^{\dagger} U^{\dagger}
\end{pmatrix} \longleftrightarrow 
\begin{pmatrix}
U_s a^{\dagger}_{\uu} U_s^{\dagger} \\
U_s a^{\dagger}_{\dd} U_s^{\dagger}
\end{pmatrix}=
\mathcal{M}^T 
\begin{pmatrix}
a^{\dagger}_{\uu} \\
a^{\dagger}_{\dd}
\end{pmatrix}.
\end{align}
We can find a matrix representation for $U_s$ by choosing a representation for the spin eigenstates
\begin{align} \tag{A.3}
\ket{\uu} \equiv
\begin{pmatrix}
1 \\
0
\end{pmatrix}, \, \, \, \, 
\ket{\dd} \equiv
\begin{pmatrix}
0 \\
1
\end{pmatrix}.
\end{align}
Inserting this representation into the relations $U_s \ket{\uu}=\alpha \ket{\uu}+\beta \ket{\dd}$ and $U_s \ket{\dd}= \gamma \ket{\uu}+\delta \ket{\dd}$, we can see that a $2 \times 2$ matrix representation of $U_s$ when it acts on the spin states is equivalent to the matrix $\mathcal{M}$. We can now see the correspondence between the evolution of the photonic two-mode state and the evolution of the spin state
\begin{align} \tag{A.4}
U\ket{M, N-M} \longleftrightarrow U_s^{\otimes N} \ket{\xi_0}_{\text{spin}},
\end{align}
where a matrix representation of $U_s$ is equivalent to $\mathcal{M}$.
\section{Proof that the Holland-Burnett (NOON) state is the only optimal state when $N=2$ ($N=3$)}
The optimality condition for photonic states can be written as a constraint on the two-particle reduced density matrix of its analogous spin state. This constraint can be expressed in the spin state form in Eq.~\eqref{eq:rhoz2new} and in the photonic form in Eq.~\eqref{eq:rho2zoptimal}. For an $N=2$ state, the two-particle reduced density matrix is simply the matrix itself. This means that a simple inspection of Eq.~\eqref{eq:rho2zoptimal} reveals the Holland-Burnett state as the only pure state satisfying this constraint.\\

A general $N=3$ photonic state takes the form $|\Psi\rangle=c_0 |0,3\rangle+c_1 |1,2\rangle+c_2|2,1\rangle+c_3|3,0\rangle$. This means that its equivalent spin state can be written as $\ket{\xi_0}_{\text{spin}}=c_0\ket{\downarrow \downarrow \downarrow}+(c_1/\sqrt{3})(\ket{\uparrow \downarrow \downarrow}+\ket{\downarrow \uparrow \downarrow}+\ket{\downarrow \downarrow \uparrow})+(c_2/\sqrt{3})(\ket{\uparrow \uparrow \downarrow}+\ket{\uparrow \downarrow \uparrow}+\ket{\downarrow \uparrow \uparrow})+c_3 \ket{\uparrow \uparrow \uparrow}$, which must satisfy Eq.~\eqref{eq:rhoz2new} for optimal states. We can rewrite Eq.~\eqref{eq:rhoz2new} in matrix form as
\begin{align} \label{eq:rhoz2matrix} \tag{B.1}
\rho^{[2]}_z=\begin{pmatrix}
\frac{1}{4}+c_{zz} & 0 & 0 & 0 \\
0 & \frac{1}{4}-c_{zz} & \frac{1}{4}-c_{zz} & 0 \\
0 & \frac{1}{4}-c_{zz} & \frac{1}{4}-c_{zz} & 0 \\
0 & 0 & 0 & \frac{1}{4}+c_{zz}
\end{pmatrix}.
\end{align}
The two-particle reduced density matrix $\text{tr}_{[2]} (\ket{\xi_0}\bra{\xi_0}_{\text{spin}})$ for $\ket{\xi_0}_{\text{spin}}$ can be shown to be
\begin{align} \label{eq:rho2zmatrix2} \tag{B.2}
\text{tr}_{[2]} (\ket{\xi_0}\bra{\xi_0}_{\text{spin}})=
\begin{pmatrix}
|c_3|^2+\frac{|c_2|^2}{3} & \frac{c_3 c_2^*}{\sqrt{3}}+\frac{c_1^* c_2}{3} & \frac{c_3 c_2^*}{\sqrt{3}}+\frac{c_1^* c_2}{3} & \frac{c_3 c_1^*}{\sqrt{3}}+\frac{c_2 c_0^*}{\sqrt{3}} \\
\frac{c_1 c_2^*}{3}+\frac{c_2 c_3^*}{\sqrt{3}} & \frac{|c_1|^2+|c_2|^2}{3} & \frac{|c_1|^2+|c_2|^2}{3} & \frac{c_1 c_0^*}{\sqrt{3}}+\frac{c_1^* c_2}{3} \\
\frac{c_1 c_2^*}{3}+\frac{c_2 c_3^*}{\sqrt{3}} & \frac{|c_1|^2+|c_2|^2}{3} & \frac{|c_1|^2+|c_2|^2}{3} & \frac{c_1 c_0^*}{\sqrt{3}}+\frac{c_1^* c_2}{3} \\
\frac{c_0 c_2^*}{\sqrt{3}}+\frac{c_1 c_3^*}{\sqrt{3}} & \frac{c_0 c_1^*}{\sqrt{3}}+\frac{c_1 c_2^*}{3} & \frac{c_0 c_1^*}{\sqrt{3}}+\frac{c_1 c_2^*}{3} & |c_0|^2+\frac{|c_1|^2}{3}
\end{pmatrix}. 
\end{align}
It is then straightforward to show that $\text{tr}_{[2]} (\ket{\xi_0}\bra{\xi_0}_{\text{spin}})$ only satisfies Eq.~\eqref{eq:rhoz2matrix} when $c_1=0=c_2$. Note that while the cases (i) $c_0=0=c_2=c_3$; (ii) $c_0=0=c_1=c_3$; (iii) $c_1=0=c_2=c_3$ in Eq.~\eqref{eq:rho2zmatrix2} obeys Eq.~\eqref{eq:rhoz2matrix} for the zero matrix elements of $\rho^{[2]}_z$, they do not obey the diagonal terms of the matrix $\rho^{[2]}_z$ in Eq.~\eqref{eq:rhoz2matrix}. The $c_1=0=c_2$ condition corresponds to the $N=3$ NOON state $\ket{\Psi}=(1/\sqrt{2})(\ket{3,0}+\ket{0,3})$.
\label{app:noon3optimal}
\end{appendices}
\section*{References}

\bibliographystyle{iopart-num}
\bibliography{RevisedQST100115.bib}
\end{document}